\documentclass[12pt]{article}

\usepackage{amsmath}
\usepackage{graphicx}
\usepackage{amssymb}
\usepackage{color}
\usepackage{bm}
\usepackage{authblk}

\usepackage[labelformat=parens]{subfig}
\usepackage{caption}

\usepackage[inline]{enumitem}

\definecolor{light-gray}{gray}{0.7}

\begin{document}

\title{A pore-scale study of transport of inertial particles by water in porous media}

\author[1]{M. A. Endo Kokubun\footnote{email: max.kokubun@uib.no}}
\author[4]{A. Muntean\footnote{email: adrian.muntean@kau.se}}
\author[2]{F.A. Radu\footnote{email: florin.radu@uib.no}}
\author[4]{K. Kumar\footnote{email: kundan.kumar@kau.se}}
\author[3]{I.S. Pop\footnote{email: sorin.pop@uhasselt.be}}
\author[2]{E. Keilegavlen\footnote{email: eirik.keilegavlen@uib.no}}
\author[1]{K. Spildo\footnote{email: kristine.spildo@uib.no}}

\affil[1]{\small{\it Department of Chemistry, University of Bergen, Norway}}
\affil[2]{\small{\it Department of Mathematics, University of Bergen, Norway}}
\affil[3]{\small{\it Faculty of Science, University of Hasselt, Belgium}}
\affil[4]{\small{\it Department of Mathematics and Computer Science, University of Karlstad, Sweden}}

\date{}

\maketitle

\begin{abstract}
We study the transport of inertial particles in water flow in porous media.
 {
Our interest lies in understanding the accumulation of particles including the possibility of clogging.
We propose that accumulation can be a result of hydrodynamic effects: the tortuous paths of the porous medium generate regions of dominating strain/vorticity, which favour the accumulation/dispersion of the inertial particles.
}
Numerical simulations show that essentially two accumulation regimes are identified: for low and for high flow velocities.
 {When particles accumulate in high-velocity regions, at the entrance of a pore throat, a clog is formed.
The formation of a clog significantly modifies the flow, as the partial blockage of the pore causes a local redistribution of pressure.}
This redistribution can divert the upstream water flow into neighbouring pores.
Moreover, we show that accumulation in high velocity regions occurs in heterogeneous media, but not in homogeneous media, where we refer to homogeneity with respect to the distribution of the pore throat diameters.
\end{abstract}

\section{Introduction}

\begin{table}[h!]
	\centering
	\caption{Nomenclature}
	\label{symbols}
	\begin{tabular}{|l |l |l |l| l|}
	\hline\noalign{\smallskip}
	$d_p$               & particle diameter               & $\alpha,\beta$      & constants of the Gumbel distribution \\
	$l$                 & characteristic pore length      &	$\eta,\sigma$       & constants of the bimodal fit \\	
    $m$                 & mixture probability             &	$\lambda$           & dimensionless value of edge length\\
	$p$                 & pressure                        &	$\mu$               & viscosity of water \\
	$t$                 & time                            &	$\xi$               & dimensionless pore throat diameter\\
	$\bm{u}$            & velocity of water               &	$\rho$              & mass density  \\	
	$\bm{v}$            & velocity of particles           &	$\tau$              & stress tensor \\ 	
	$x,y$               & spatial coordinates             &	$\varphi$           & volumetric occupation of particles \\	
\noalign{\smallskip}\hline
	\end{tabular}
\end{table}

The transport of particles in porous media emerge in many problems of interest, such as spillage of contaminants in soils \cite{mccarthy1989,keller2007}, water filtration systems \cite{rezakazemi2018}, fines migration \cite{russel2018,chequer2019}, enhanced oil recovery \cite{mack1994,bedrikovetsky2017}, to name a few.
For enhanced oil recovery methods (EOR), injection of nanoparticles along with water has shown to recover initially trapped oil \cite{li2011,elamin2013,bedrikovetsky2016}.
The enhancement can be an effect of a favourable change on the relative permeability curves \cite{bedrikovetsky2016} or improved mobility ratio \cite{li2011ef}, for example.
%
%
 {
For the particular case when the nanoparticles are polymer particles, successful field trials, with increased oil recovery and lower water production were conducted for reservoirs with highly heterogeneous permeabilities \cite{mack1994}.
However, the driving mechanism behind this EOR technique is still unclear.
}

 {
To clarify the underlying physical mechanism of EOR through polymer particles injection, experiments at the laboratory scale were performed using reservoir sandstone cores.
Results from coreflood experiments show that the enhancement in oil recovery is significantly stronger in porous media with a heterogeneous pore-size distribution, compared to recovery in microscopically homogeneous media \cite{spildo2009}.
Moreover, for polymer particles in sandstone cores, particles and rock have the same surface charge \cite{skauge2008,nasralla2014}.
Deposition due to electrostatic attraction is thus unlikely.
Based on these experimental findings, a mechanism of log-jamming and microscopic flow diversion was proposed as responsible for EOR \cite{spildo2009,spildo2010}.
This effect is believed to take place when particle-carrying water flows from a large to a narrow pore channel. 
The water then accelerates, and particles are left behind at the entrance of the pore due to inertia. 
Particles may thus accumulate, eventually leading to clogging of the pore, and diversion of the upstream flow into neighbouring channels. 
If the flow is diverted into a channel which contains that was trapped in the initial stages of waterflood, the water flow can mobilize this oil.
Porous media with heterogeneous distributions of pore throat radii present high frequency of transitions from large to narrow pores compared to homogeneous media. 
Hence log-jamming and microscopic flow diversion as governing processes is consistent with the observed enhancement in recovery in heterogeneous media.
}

 {
To further probe the hypothesized mechanism, the flow process can be studied by numerical simulations. 
These can be carried out either on the core scale, where comparison with the laboratory measurements are feasible, or on finer scales where the flow dynamics can be represented in more detail.
On the core scale, simulations of oil-water flow in porous media with injection of polymer particles were carried out by Kokubun {\it et al.} \cite{kokubun2019}, with log-jamming represented by an ad hoc non-equilibrium reaction rate that accounted for core heterogeneity
The results showed that the simple core-scale model is able to qualitatively reproduce the oil recovery and pressure drop across the core observed in laboratory experiments.
However, the lack of a proper model for the entrapment rate of particles through log-jamming made it difficult to conclude on the fundamental driving mechanism behind increased recovery.
A finer scale approach for analysing pore-scale physics of multiphase flow in porous media is through the use of pore network models \cite{blunt2002}.
This method is able to extract useful relations, such as capillary pressure and relative permeabilities \cite{jivkov2016}.
Still the representation of flow within pore network models is simplified compared to the real pore-scale flow.
Simulations of polymer particle transport, for example, rely on postulated entrapment rates, e.g. through a condition derived from breakthrough curves \cite{bolandtaba2009}.
Moreover, the pore network model for polymer particles transport developed by Bolandtaba {\it et al.} \cite{bolandtaba2009} does not take into account the difference in velocities of water and particles at the entrance of the narrow pore.
}

 {
In this work, we aim to perform a preliminarily investigation of the log-jamming mechanism by modeling from first principles, based on a computational fluid dynamics (CFD)  approach.
The starting point to construct a proper mathematical model is to identify the relevant physical mechanisms that must be taken into account, while simultaneously trying to limit the model complexity to simplify later numerical treatment.
With this in mind, to model accumulation of particles due to inertia, water and particles should be allowed to have different velocities.
Conversely, the injection of polymer particles along with water in EOR usually occurs when the oil is no longer mobilized by water alone.
Thus, the initial dynamics of particles transport and accumulation in narrow channels will happen in water-accessible channels. 
That is, the process is locally a water flow regime, and we therefore chose to neglect the oil phase in our model, and include only the flow of water with particles.
}

 {
This paper has three main contributions relevant to the study of particle accumulation, clogging and flow diversion. 
First, based on the above assumptions we derive a multiphase flow model that considers the particles as a dispersed phase in the water phase, using an Euler-Euler formulation.
Secondly, by numerical simulations of the model in random media we show that the derived model has a tendency to produce particle accumulation, clogging and flow diversion at the transition from wide to narrow pores.
This effect is however naturally dependent not only on geometry, but also on dynamic factors such as the local flow direction and velocity.
The third contribution is therefore to develop an indicator, based on water velocity strain rates, that identifies regions in the pore-space where accumulation takes place. 
This measure is computed from simulations of  water flow only, with no particles included, and is thus computationally relatively cheap.
The observed correlation between regions of strain and particle accumulation, and conversely vorticity and particle dispersion, is consistent with previous observations in the transport of inertial particles by turbulent flows \cite{fessler1994,falkovich2004,boffetta2007}.
In non-confined flow, the cause of regions of large strain or vorticity, and consequently particle accumulation and dispersion, is turbulence. 
Conversely, while flows in porous media is well below the turbulent regime, the flow is still highly heterogeneous on the pore-scale due to the tortuous paths of the porous medium \cite{deanna2017}.
This creates regions of dominating strain and vorticity even in the laminar regime.
We identify distinct accumulation scenarios, not all leading to the formation of clogs, but all related to strain-dominated regions.
We emphasise on the fact that the clogging is understood here as the result of particle accumulation, where the particles take a non-negligible part of the fluid. 
We do not account for other clogging mechanisms, like precipitation or dissolution/adsorption \cite{bringedal2017,noorden2009}, or the growth of biofilm \cite{landa2019,schulz2019}. 
}

We begin by introducing the multiphase flow model and present its dimensionless formulation.
For the numerical simulations, we consider porous media consisting of non-overlapping solid circles distributed in a $2D$ rectangular box.
We evaluate the influence of the parameters of interest: Stokes number, Reynolds number and particle-to-water mass densities ratio.
Also, we show that the accumulation pattern is distinct if we consider a heterogeneous or a homogeneous media.
In particular, the initial formation of clogs (accumulation in narrow pore throats that may lead to flow diversion) only occurs in heterogeneous media.

\section{Model formulation}

We follow a general derivation based on the classical mixture theory.
 {
For details on the derivation of the model, see the Appendix.
}
We consider that the particulate phase is dispersed in the water phase, and hence, it can be described by a continuous field.
The fractional volumetric occupation of the particles is given by $\varphi\in [0,1]$, whereas the water phase occupies a volumetric fraction of $1-\varphi$.
We assume that the only internal interaction between the dispersed and the 
water phase is the Stokes drag.
Additionally, if we model the mass densities of water and particles, $\rho$ and 
$\rho_p$, respectively, as constant, we have the following set of governing 
equations
\begin{align}
\frac{\partial(1-\varphi)}{\partial t}
+
\nabla\cdot((1-\varphi)\bm{u})
&=
0,
\label{eq.massW}
\\[5pt]
\frac{\partial\varphi}{\partial t}
+
\nabla\cdot(\varphi\bm{v})
&=
0,
\label{eq.massP}
\\[5pt]
\rho(1-\varphi)
\frac{D\bm{u}}{Dt}
&=
-
(1-\varphi)
\nabla p
+
\nabla\cdot
\left(
(1-\varphi)\bm{\tau}_w
\right)
-
\frac{\varphi\rho_p}{t_s}(\bm{u}-\bm{v}),
\label{eq.momW}
\\[5pt]
\rho_p\varphi
\frac{d\bm{v}}{dt}
&=
-
\varphi\nabla p
+
\frac{\varphi\rho_p}{t_s}(\bm{u}-\bm{v}),
\label{eq.momP}
\end{align}
where $\bm{u}$ and $\bm{v}$ are the velocities of water and particles, respectively
and $t_s = \rho_p d_p^2/(18\mu)$ is the Stokes time, or the characteristic 
response time of the particles with respect to changes in the flow field, with 
$d_p$ the particle diameter \cite{kleinstreuerBook}.
The stress tensor $\bm{\tau}_w$ of the water phase is given by
\begin{equation}
\bm{\tau}_w
=
\mu
\left(
\nabla\bm{u} + \nabla\bm{u}^T
\right)
-
\frac{2}{3}
\mu\bm{I}\nabla\cdot\bm{u},
\label{eq.tauW}
\end{equation}
where $\mu$ is the dynamic viscosity of pure water.
Moreover, we use the notation
\begin{equation}
\frac{D}{Dt} = \frac{\partial}{\partial t} + \bm{u}\cdot\nabla, \ \ \
\frac{d}{dt} = \frac{\partial}{\partial t} + \bm{v}\cdot\nabla,
\label{eq.ddt}
\end{equation}
which are the material derivatives along the water and particle streamlines, 
respectively.
Boundary and initial conditions are added further ahead to complete the model.

The model given by Eqs. (\ref{eq.massW})--(\ref{eq.momP}) assumes that the 
particles are small enough not to disturb the water flow locally by forming 
vortices, and potentially turbulence, behind them.
Formally, we assume that the particles diameter is such that the flow around 
the particles is in the Stokes regime.
That is, the particle-based Reynolds number, $Re_p = \rho u_c d_p/\mu$ is less 
than unity, where $u_c$ is a characteristic flow velocity.
The flow velocities in porous media exhibit a variation of several orders of 
magnitude, typically from $1-100~\mu \mbox{m}/\mbox{s}$ \cite{kutsovsky1996}.
We are interested in inertial particles in the colloidal scale, i.e., $d_p\sim10^{-4}\mbox{m}$, such that for water properties evaluated at $300K$, we have $Re_p \sim 10^{-6}-10^{-4}$, typically.
Hence, the assumption of Stokes flow around the particles is justified
for these conditions. We emphasize that the background flow does not have such 
restrictions.

The no-slip flow condition is considered for the water at the surface of the solid grains forming the porous medium.
The boundary conditions of the particles velocity at the wall are more 
intricate, as the no-slip and no-flow conditions are not necessary true for 
particulate flow.
For example, a perfect elastic collision between particle and wall implies 
specular reflection, i.e., $[\bm{v}\cdot\bm{n}] = -2(\bm{v}\cdot\bm{n})\bm{n}$ 
and $[\bm{v}\cdot\bm{t}] = 0$, where $\bm{n}$ and $\bm{t}$ are the unitary 
vectors normal and tangential to the solid boundaries, respectively.
We consider the limit of asymptotically small Stokes number, such that an explicit expression for $\bm{v}$ is used instead of Eq. (\ref{eq.momP}).
In this case, the particles velocity at the wall is prescribed.
The error in the calculation of the particles velocity at the wall introduced by this approximation is only relevant close to the wall.

\subsection{Non-dimensionalization}

We consider the following non-dimensional variables for a $2D$ geometry in the $(x,y)$ Cartesian coordinates
\[
\hat{t}   = t/t_c, \ \ \
\hat{x} = x/l, \ \ \
\hat{y} = y/l, \ \ \
\]
\begin{equation}
\hat{\bm{u}} = \bm{u}/u_c, \ \ \
\hat{\bm{v}} = \bm{v}/u_c, \ \ \
\hat{p}      = p/p_c ,
\label{eq.non}
\end{equation}
where $u_c$ and $l$ are characteristic values for velocity and  {pore length}, with the characteristic time given by $t_c = l/u_c$, and $p_c= \mu u_c/l$ a characteristic pressure.
The non-dimensional governing equations are then given by (we remove the hats for simplicity)
\begin{align}
\nabla\cdot
\left(
(1-\varphi)\bm{u} + \varphi\bm{v}
\right)
&=
0,
\label{eqND.mass}
\\[5pt]
\frac{\partial\varphi}{\partial t}
+
\nabla\cdot(\varphi\bm{v})
&=
0,
\label{eqND.massP}
\end{align}
\begin{align}
(1-\varphi)
\frac{D\bm{u}}{Dt}
&=
-
\frac{1}{Re}
(1-\varphi)\nabla p
+
\frac{1}{Re}
\nabla\cdot
\left(
(1-\varphi)
\left(
\nabla\bm{u} + \nabla\bm{u}^T
-
\frac{2}{3}
\bm{I}\nabla\cdot\bm{u}
\right)
\right)
-
\frac{\varphi\tilde{\rho}}{St}(\bm{u}-\bm{v}),
\label{eqND.momW}
\\[5pt]
\tilde{\rho}
\frac{d\bm{v}}{dt}
&=
-
\frac{1}{Re}\nabla p
+
\frac{\tilde{\rho}}{St}(\bm{u}-\bm{v}),
\label{eqND.momP}
\end{align}
where Eq. (\ref{eqND.mass}) is obtained by adding the dimensionless form of Eqs. (\ref{eq.massW}) and (\ref{eq.massP}) and we introduce the following dimensionless numbers
\begin{equation}
Re             = \frac{\rho u_c l}{\mu}, \ \ \
St             = \frac{t_s}{t_c}, \ \ \
\tilde{{\rho}} = \frac{\rho_p}{\rho},
\label{eqND.numbers}
\end{equation}
which are the Reynolds number, the Stokes number and the ratio of pure mass densities, respectively.
 {
For typical values of $u_c\approx10^{-4}~\mbox{m}/\mbox{s}, l \approx 10^{-3}~\mbox{m}, \rho \approx 10^3~\mbox{kg}/\mbox{m}^3$ and $\mu \approx 10^{-3}~\mbox{Pa}\cdot\mbox{s}$, we have $Re \sim O(1)$ and $St \ll 1$.
For the mass densities ratio, we consider particles heavier than water, i.e., $\tilde{\rho} > 1$.
}

As mentioned previously, the model is limited by the constraint of a Stokes flow regime around the particles, i.e., $Re_p = \rho u_c d_p/\mu = Re (d_p/l) < 1$.
Therefore, the Stokes regime around the particles is obeyed for $d_p/l < Re^{-1}$.
Since $t_s = \rho_p d_p^2/(18\mu)$, we can write the Stokes number as $St = \tilde{\rho} (Re/18) (d_p/l)^2$.
Then, we write the constraint of Stokes regime around the spheres as
\begin{equation}
 St < \frac{\tilde{\rho}}{18 Re}.
 \label{eq.St}
\end{equation}

\subsection{Asymptotic limit of $St \ll 1$}

 {
We can simplify the momentum equation for the particles, Eq. (\ref{eqND.momP}), by considering the asymptotic limit of small Stokes numbers, i.e., $St\ll1$.
The objective of such simplification is two-fold: first, the asymptotic limit provides an explicit expression for $\bm{v}$, minimizing computational costs, and second, it provides a framework for quantifying preferential accumulation/dispersion regions, as will be shown further.
}
This limit represents a small response time of the particles with respect to changes in the surrounding flow field.
In other words, the inertia effects are limited.

In this limit, we expand the particles velocity in powers of $St$ as
\begin{equation}
 \bm{v} = \bm{v}_0 + St\bm{v}_1 + o(St),
 \label{eqAE.v}
\end{equation}
where $o(St)$ is interpreted as a Landau symbol.
Using (\ref{eqAE.v}) into Eq. (\ref{eqND.momP}), which yields $d/dt = D/Dt + o(1)$, and equating the terms of order $0$ and $1$ in $St$ gives the following asymptotic expression for $\bm{v}$
\begin{equation}
\bm{v} = \bm{u} - St\left(\frac{1}{\tilde{\rho}Re}\nabla p + \frac{\partial \bm{u}}{\partial t} + \bm{u}\cdot\nabla\bm{u}\right) + o(St),
\label{eq.v}
\end{equation}
where it is required that $\tilde{\rho}Re > o(St)$.
Equation (\ref{eq.v}) is a modified version of the expression first obtained by Maxey \cite{maxey1987} and used recently to study particle-laden turbulent flows \cite{boffetta2007,falkovich2004,mitra2018,moghadam2016}.
In the present case, we consider an additional contribution due to the pressure, and this difference is important for the near-wall treatment of the particles velocity.
Moreover, using $d/dt$ instead of $D/Dt$, that is, material derivatives along 
$\bm{v}$ instead of $\bm{u}$ allows for an explicit expression for $\bm{v}$, 
which facilitates the problem analysis, as shown ahead.

Using (\ref{eq.v}) up to the order of $St$ we have the following system of governing equations
\begin{align}
\nabla\cdot
\left(
(1-\varphi)\bm{u} + \varphi\bm{v}
\right)
&=
0,
\label{eqAL.mass}
\\[5pt]
\frac{\partial\varphi}{\partial t}
+
\nabla\cdot(\varphi\bm{v})
&=
0,
\label{eqAL.massP}
\end{align}
\begin{align}
(1-\varphi)
\frac{D\bm{u}}{Dt}
&=
-
\frac{1}{Re}
(1-\varphi)\nabla p
+
\frac{1}{Re}
\nabla\cdot
\left(
(1-\varphi)
\left(
\nabla\bm{u} + \nabla\bm{u}^T
-
\frac{2}{3}
\bm{I}\nabla\cdot\bm{u}
\right)
\right)
-
\frac{\varphi\tilde{\rho}}{St}(\bm{u}-\bm{v}),
\label{eqAL.momW}
\\[5pt]
\bm{v}
&=
\bm{u} - St\left(\frac{1}{\tilde{\rho}Re}\nabla p + \frac{\partial \bm{u}}{\partial t} + \bm{u}\cdot\nabla\bm{u}\right).
\label{eqAL.momP}
\end{align}

\noindent
We can rewrite Eq. (\ref{eqAL.massP}) as
\begin{equation}
\frac{1}{\varphi}
\frac{d\varphi}{dt}
=
- \nabla\cdot\bm{v}.
\label{eqAL.massPb}
\end{equation}
Therefore, accumulation or dispersion of particles along its streamlines is characterized by the quantity
\begin{equation}
\nabla\cdot\bm{v}
=
\nabla\cdot\bm{u}
-
St\left(
\frac{1}{\tilde{\rho}Re}\nabla^2 p + \frac{\partial}{\partial t}\nabla\cdot\bm{u}
+
\nabla\cdot(\bm{u}\cdot\nabla\bm{u})
\right).
\label{eqAL.divV}
\end{equation}
According to Eqs. (\ref{eqAL.massPb}) and (\ref{eqAL.divV}), when $\nabla\cdot\bm{v}<0$, accumulation is favoured, whereas dispersion is favoured for $\nabla\cdot\bm{v}>0$.
Note that $\nabla\cdot\bm{u}\neq0$ may happen in certain areas of the domain due to the inherent compressibility, see Eq. (\ref{eqAL.mass}).

We consider no-slip flow for the water at the solid boundaries, i.e., $|\bm{u}| = 0$, at these boundaries.
Therefore, Eq. (\ref{eqAL.momP}) determines $\bm{v} = - St(\tilde{\rho}Re)^{-1}\nabla p$ as the particles velocity at solid boundaries.
This condition implies slip flow and reflection of particles at the wall.
The error introduced depends on the original boundary condition for the 
particles.
For example, if we had specular reflection, the slip flow at the wall is unrealistic, but the reflection is not.
Nevertheless, the error is restricted to the vicinity of solid boundaries.
We will show that due to vorticity these regions are characterized by dispersion of particles, such that this error is not expected to be relevant.

\section{Numerical results}

\begin{figure}[h!]
\centering
\includegraphics[width=1\linewidth]{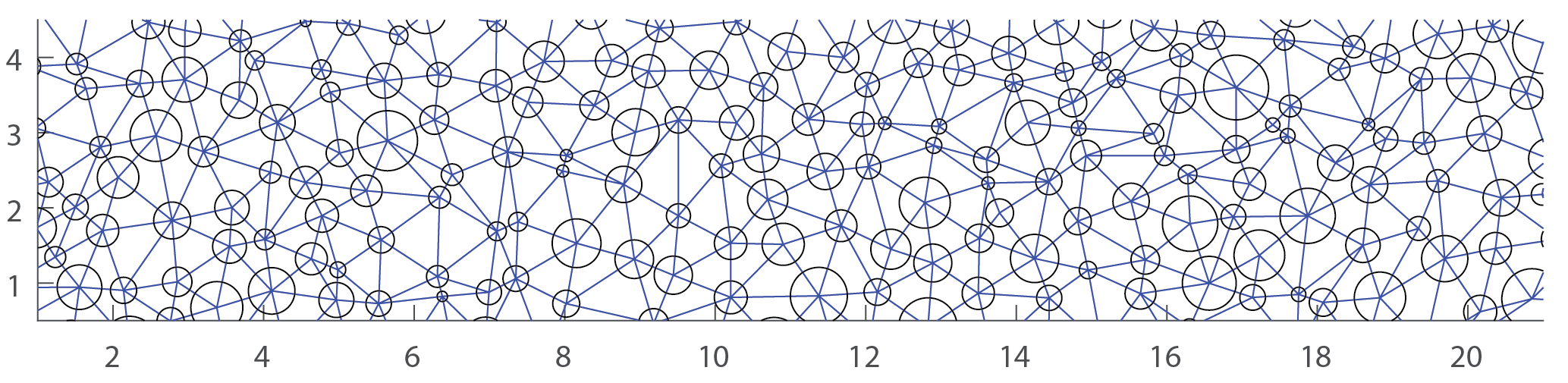}
\caption{Illustrative porous medium. The nodes of the Delaunay triangulation provide the coordinates of the centres of the solid grains as well as the basis for calculating the pore throat diameters. The numerical domain comprises the void space between the solid grains (the mesh is not shown here).  {The axis are shown in dimensionless form and the $x/y$ aspect ratio of the porous medium is $0.2$.}}
\label{fig.porScheme}
\end{figure}

 {
In this Section, the model for pore-scale transport of water and particles, 
given by Eqs. (\ref{eqAL.mass})--(\ref{eqAL.momP}), is solved in 
artificially generated two-dimensional porous media.
Motivated by the observations from core flooding experiments that EOR rates are most prominent in media with heterogeneous pore throat 
distributions, our intention is to quantify the behaviour of the model in random media with 
given pore throat diameter distributions (PTDD).
Moreover, the heterogeneous reservoir sandstone cores utilized in the experiments in \cite{spildo2009} have a bimodal distribution.
Therefore, we investigate both media with homogeneous PTDD, and media with heterogeneous PTDD which are represented by bimodal distributions. 
To generate the media, we first generate a random set of points in the simulation domain,
which we take as a rectangle; the points will become centres of solid grains. 
Nearby points are next connected by edges, where the edges are defined so as to 
form a Delaunay triangulation of the domain. 
The points are then moved around, and the connecting edges updated, to make the edge lengths meet a target 
distribution. 
Specifically, we target a sum of two Gumbel distributions \cite{lawlessBook}
}
\begin{equation}
 f_e(\lambda) = \sum_{i=1}^2\frac{1}{\beta_i}\mbox{exp}\left(-s_i(\lambda) + e^{-s_i(\lambda)}\right),
 \label{eq.gumbel}
\end{equation}
where $\lambda$ are the edge lengths and $s_i(\lambda) = (\lambda-\alpha_i)/\beta_i$, with $\alpha_i$ and $\beta_i$ given constants.
 {We target Gumbel distributions because they are highly localized, thus avoiding tail values for the edges.}
The procedure is implemented as a modified version of the distmesh algorithm for mesh generation, originally proposed in \cite{persson2004}.
Then, we draw circles centred at each node of the triangles, making sure that the circles do not overlap.
Each pore throat diameter is then given by subtracting from the length of the original triangle edge the radii of the circles located in its associated nodes.
We construct the medium such that we fit a bimodal distribution for the PTDD as a sum of two normal distributions
\begin{equation}
 f_r(\xi)
 =
 \frac{m}{\sigma_1\sqrt{2\pi}}~\mbox{exp}\left(-\frac{(\xi-\eta_1)^2}{2\sigma_1^2}\right)
 +
 \frac{1-m}{\sigma_2\sqrt{2\pi}}~\mbox{exp}\left(-\frac{(\xi-\eta_2)^2}{2\sigma_2^2}\right),
 \label{eq.fit}
\end{equation}
where $\xi = \lambda - r_1 - r_2$ is the pore throat diameter length (with $r_1$ and $r_2$ the radii of the solid grains centred at the nodes of the edges $\lambda$), $ {\eta_i}$ and $\sigma_i$ are the distribution parameters, with $i=1,2$, and $0<m<1$ the mixture probability.
For convention, $\eta_1$ and $\sigma_1$ refer to the normal distribution around the smaller pores, whereas $\eta_2$ and $\sigma_2$ refer to the larger pores.
The procedure is similar to the one described in \cite{deanna2017}.

We also define a parameter $\gamma$ as
\begin{equation}
 \gamma = \frac{(1-m)\eta_2}{m~\eta_1},
 \label{eq.gamma}
\end{equation}
which is a measurement of the large-to-narrow pore throat diameter ratio.
For $\gamma\ll1$, there is a predominance of small pores, whereas for $\gamma\gg1$ there is a predominance of large pores.
For both limiting cases $\gamma\rightarrow0$ and $\gamma\rightarrow\infty$, the medium is homogeneous,  {i.e., there is a predominance of same-size pores.}

We implement Eqs. (\ref{eqAL.mass})--(\ref{eqAL.momP}) in COMSOL, thus numerically solving them using a fully coupled, standard Galerkin finite element method.
The water velocity $\bm{u}$ is discretized using quadratic Lagrange elements, whereas the pressure $p$ and the volumetric fraction of particles $\varphi$ are discretized using linear Lagrange elements.
The system of equations is solved implicitly in a domain meshed by a Delaunay 
tesselation, see Figure \ref{fig.porScheme} for an example of the computational 
domain and mesh.
We consider a parabolic injection profile for the water as $\bm{u} = (4u_{inj}(y-y_{min})(y_{max}-y)/(y_{max}-y_{min})^2,0)$.
The numerical domain will extend from $x\in[-1,26]$ and $y\in[0.5,4.5]$, but the solid grains fill the box $(1\leq x\leq21)\cup(0.5\leq y\leq4.5)$.
 {
Thus, $x/y = 0.2$ is the aspect ratio of our porous domain.
Considering a typical value of pore length of $l\sim10^{-3}~m$, we are considering a porous domain of the order of $10^{-2}~m$.
}

For numerical stability, we introduce an artificial diffusion term in Eq. (\ref{eqAL.massP}), such that we essentially solve
\begin{equation}
\frac{\partial \varphi}{\partial t}
+
\nabla\cdot\left(
\varphi\bm{v} - \frac{1}{Pe}\nabla\varphi
\right)
=
0,
\label{eqAL.momPmod}
\end{equation}
and we consider $Pe = 20$.
It is worth noting that the artificial diffusion can be interpreted as particle diffusion in the water phase.
Moreover, since $Pe>0$, it tends to lower $\varphi$, i.e., disperse particles concentration (see Eq. (\ref{eqAL.divV})).
Nevertheless, as long as the advection transport is of unitary-order or lower, the influence of $Pe$ is very small, such that the role of diffusion is restricted to numerical stabilization.

\subsection{Particle accumulation in heterogeneous domain}

\begin{figure}[h!]
\centering
\includegraphics[width=0.65\linewidth]{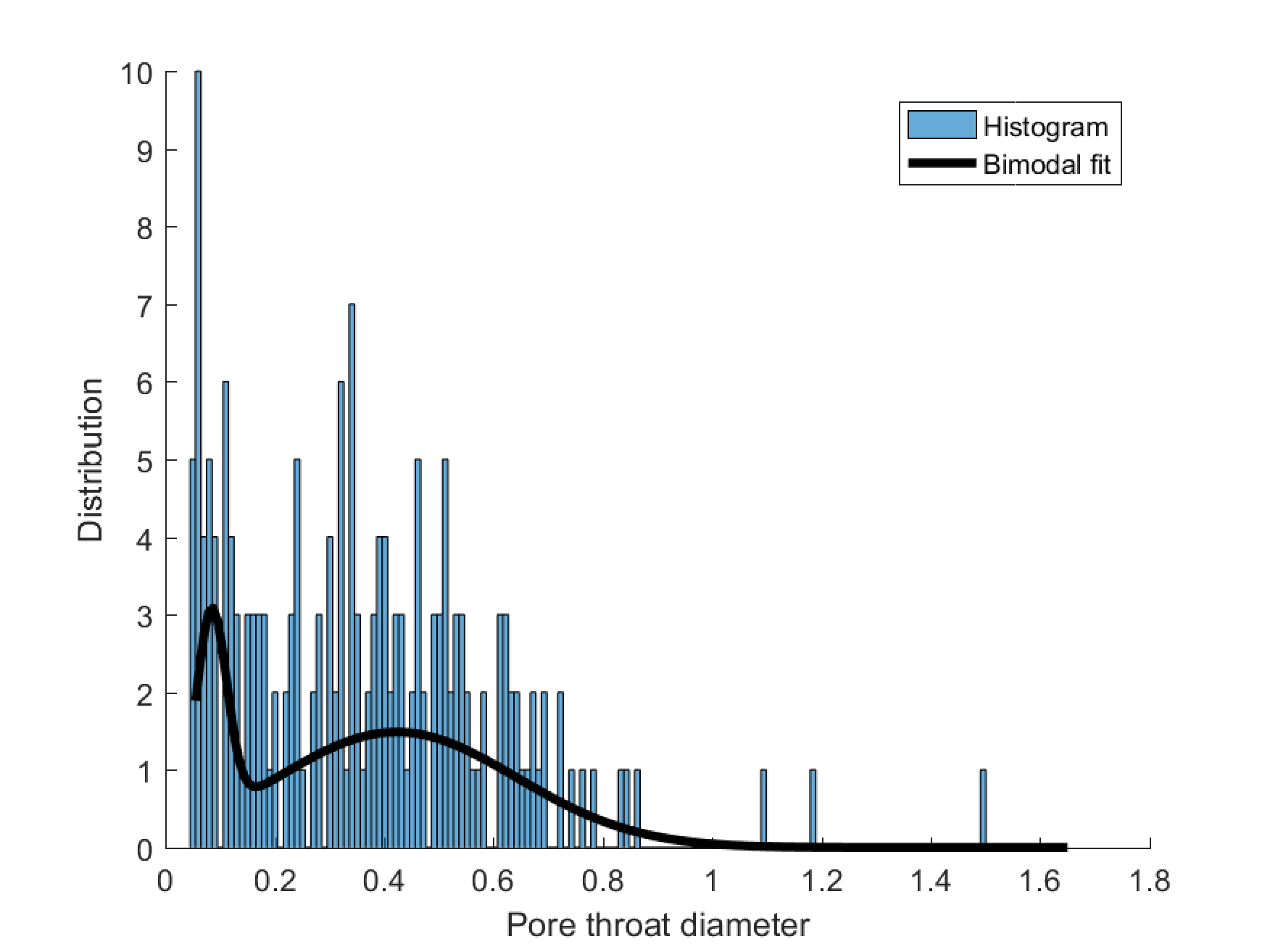}
\caption{Histogram and bimodal fit for the distribution of pore throat diameters of the medium shown in Fig. \ref{fig.porScheme}.  {The value of the diameters is given in dimensionless form.}}
\label{fig.poreSchemeB}
\end{figure}

We begin by performing numerical simulations in the medium show in Fig. \ref{fig.porScheme}.
This random medium is composed coincidently by $181$ circles and $181$ pore throats.
The parameters used for the initial edge length distribution are shown in Tab. 
\ref{tab.parGumbel}.
\begin{table}[h!]
\centering
 \begin{tabular}{|c|c|}
  \hline
  $\alpha_1 = 0.5$ & $\beta_1 = 0.05$\\
  $\alpha_2 = 0.8$ & $\beta_2 = 0.10$\\
  \hline
 \end{tabular}
 \caption{Parameters for the Gumbel distributions used to generate the Delaunay triangulation of the rectangular domain shown in Fig. \ref{fig.porScheme}.}
 \label{tab.parGumbel}
\end{table}

This medium represents a heterogeneous porous medium with a PTDD shown in Fig. \ref{fig.poreSchemeB}, where we plot the distribution of the pore throat diameters and its bimodal fit.
For the distribution shown in Fig. \ref{fig.poreSchemeB}, we have the fitting parameters shown in Tab. \ref{tab.het}.
We consider $u_{inj} = 1.2, Re = 5, St = 0.1$ and $\tilde{\rho} = 10$.
Note that according to (\ref{eq.St}) we must have $St < 0.11$ for the considered parameters.
A value of $\tilde{\rho} = 10$ represent particles with a high mass density when compared to water, silver, for example.
\begin{table}[h!]
\centering
\begin{tabular}{|c|c|}
\hline
 $p = 0.1834$         & $\gamma   = 22.3662$\\
 $\eta_1    = 0.0842$ & $\eta_2   = 0.4232$\\
 $\sigma_1 = 0.0277$  & $\sigma_2 = 0.2191$\\
\hline
\end{tabular}
\caption{Fitting parameters for the heterogeneous media shown in Fig. \ref{fig.porScheme}}
\label{tab.het}
\end{table}

\begin{figure}[h!]
\centering
\includegraphics[width=1\linewidth]{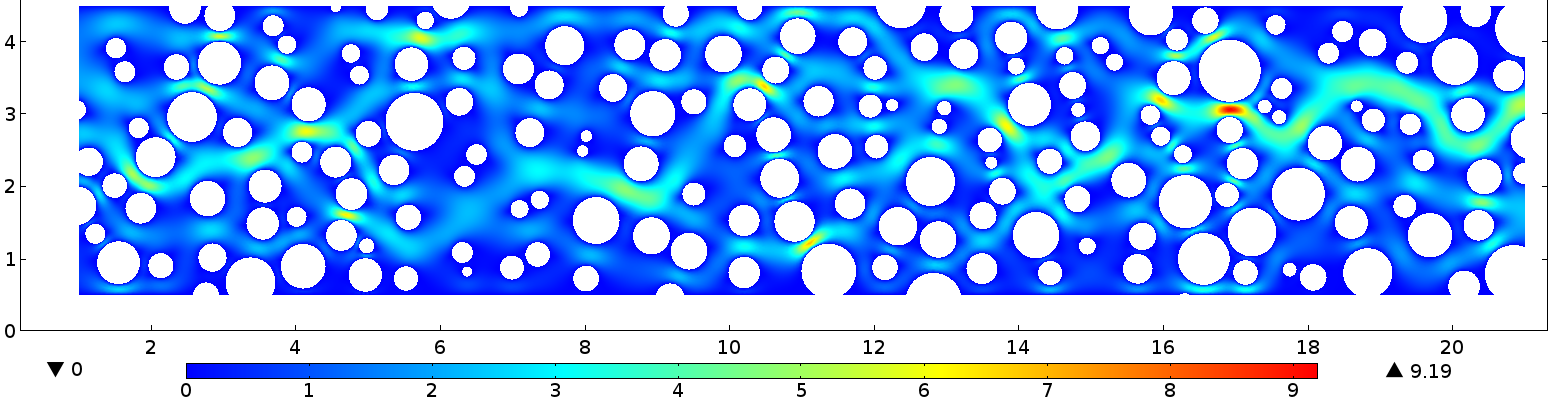}
\includegraphics[width=1\linewidth]{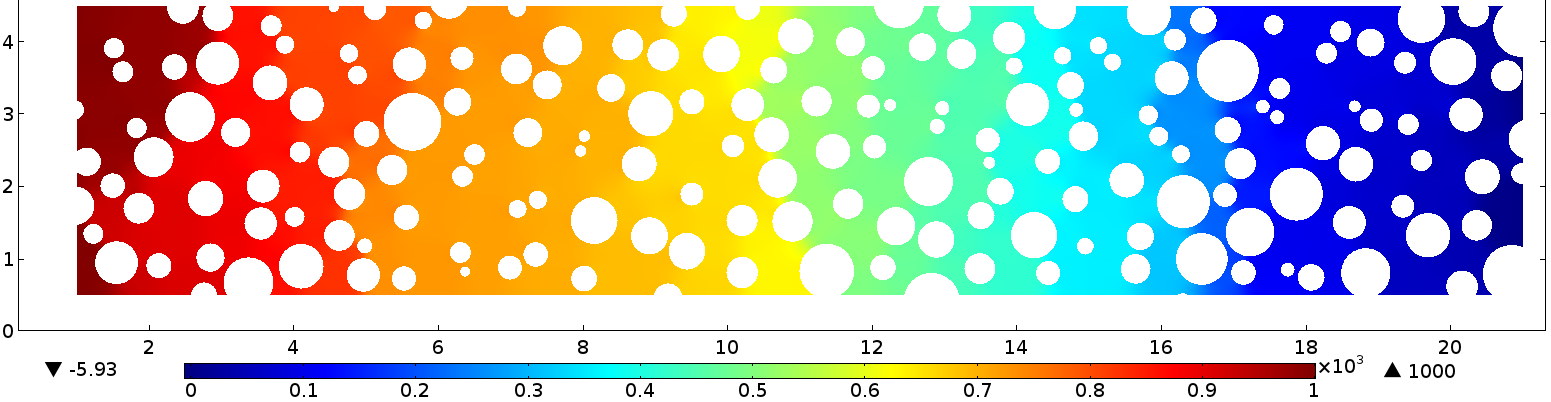}
\caption{Flow pattern without particles ($\varphi = 0$). Top: Water phase 
velocity (dimensionless). Bottom: pressure drop along the domain 
(dimensionless). The flow is from left to right and shown for $t=10$. 
 {Red colour indicates larger values of velocity and pressure, 
whereas blue indicates lower values}.}
\label{fig.waterEx}
\end{figure}

\subsubsection{Indicators of accumulation based on pure water flow}

We first consider simulations with injection of pure water, with an aim of 
studying the flow pattern, and identify regions that are likely to experience 
accumulation of particles when these are injected.
In Fig. \ref{fig.waterEx} we show a stationary water velocity field (top) and  
the corresponding pressure drop (bottom) for the case of flow without 
particles, i.e., for 
$\varphi = 0$ everywhere  {and in dimensionless form}.
The steady state was reached before $t=10$.
We see that the flow pattern is heterogeneous due to the tortuous paths of the porous medium.
This heterogeneity in the velocity field is determinant for the accumulation and dispersion of particles.
In fact, it is possible to predict regions where accumulation and dispersion are predominant.
To do so, we consider Eq. (\ref{eqAL.divV}) with $\nabla\cdot\bm{u} = 0$, which happens when $\varphi=0$,
\begin{equation}
\nabla\cdot\bm{v}
=
-
St\left(
\frac{1}{\tilde{\rho}Re}\nabla^2 p
+
\nabla\cdot(\bm{u}\cdot\nabla\bm{u})
\right).
\label{eq.divVphi0}
\end{equation}
Moreover, from the bottom Fig. \ref{fig.waterEx} one can see that the pressure drop $\nabla p$ along the domain is nearly linear, such that $\nabla^2 p \ll 1$.
Hence, we expect that, quantitatively, the first term in the right-hand side of Eq. (\ref{eq.divVphi0}) is much smaller than the second term, such that we can approximate
\begin{equation}
\nabla\cdot\bm{v}
\approx
-
St\left(
\nabla\cdot(\bm{u}\cdot\nabla\bm{u})
\right)
=
-
St
\frac{(s^2-w^2)}{2},
\label{eq.divVphi0B}
\end{equation}
where $s = \partial_y u_x + \partial_x u_y$ and $w = \partial_y u_x - \partial_x u_y$ are the local strain and vorticity of the flow, respectively.
Of course, as particles tend to accumulate, compressibility effects will take place and (\ref{eq.divVphi0B}) must be replaced by (\ref{eqAL.divV}).
Nevertheless, possible locations of accumulation and dispersion of particles can be estimated by analyzing (\ref{eq.divVphi0B}) in the domain \cite{boffetta2007,dodin2004,druzhinin1995,maxey1990}.
 {It is worth to note that for $3$D media we have $\delta = -(|\bm{S}|^2 - |\bm{W}|^2)/2$, where $|\bm{S}|$ and $|\bm{W}|$ are the modulus of the local strain and vorticity.}

\begin{figure}[h!]
\centering
\includegraphics[width=1\linewidth]{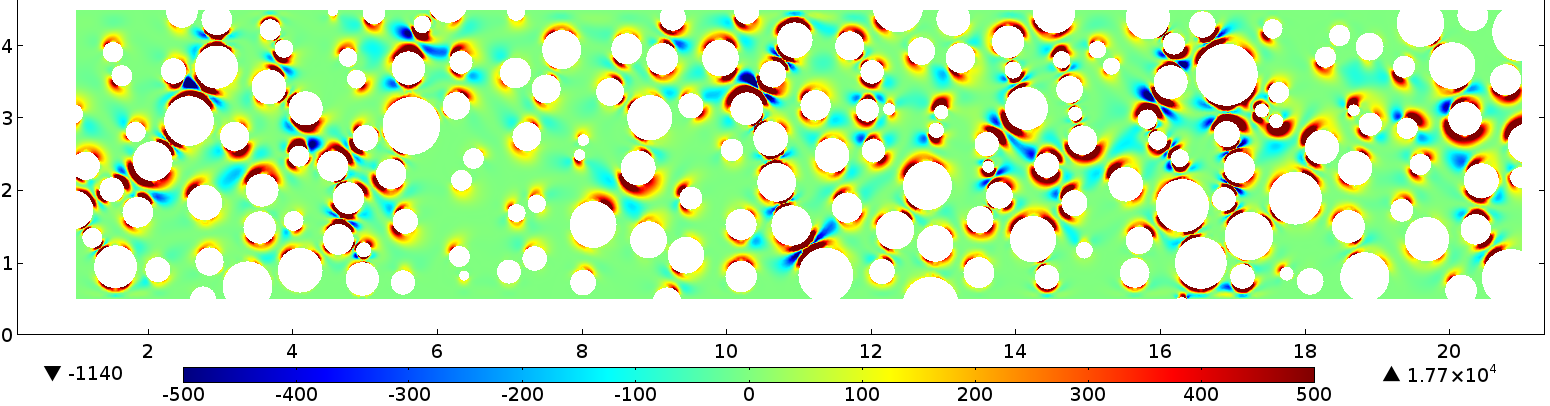}
\includegraphics[width=1\linewidth]{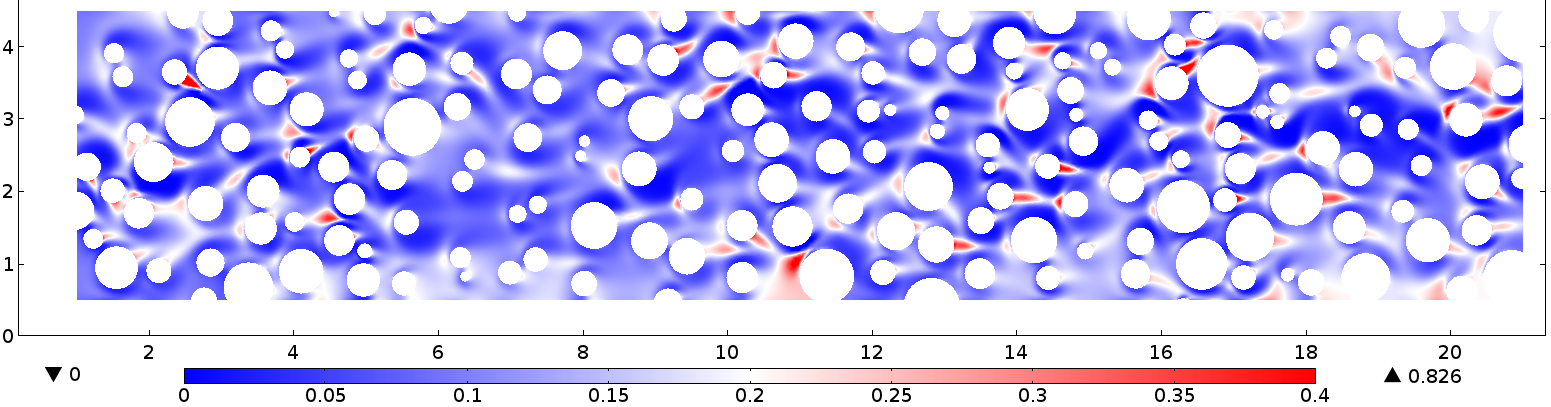}
\caption{Top: surface plot of $\delta$. Regions where $\delta<0$ (blue colour) 
favour accumulation of particles, whereas regions where $\delta>0$ (red colour) 
favour dispersion. Bottom: particle concentration for injection at 
$\varphi=0.1$. Accumulation is shown in red, whereas dispersion is shown in 
blue. Plots are shown for $t=80$ (stationary regime).}
\label{fig.divV}
\end{figure}

 {We define}
\begin{equation}
\delta = \frac{1}{St}\nabla\cdot\bm{v}
=
-
\frac{(s^2-w^2)}{2},
\label{eq.delta}
\end{equation}
 {and plot this quantity in the top of Fig. \ref{fig.divV}.
For strain-dominated regions, $\delta<0$, whereas $\delta>0$ for vorticity-dominated regions.
}
One can see that vorticity is induced near the solid boundaries of the circular grains, making particles disperse from it (positive, red regions in the top Fig. \ref{fig.divV}).
On the other hand, regions with dominating strain favour accumulation, according to (\ref{eq.divVphi0B}) and (\ref{eqAL.massPb}) (negative, blue regions in the top Fig. \ref{fig.divV}).

\subsubsection{Spatial distribution of particle accumulation}

\begin{figure}[h!]
\centering
  \subfloat[]{\includegraphics[width=0.5\linewidth]{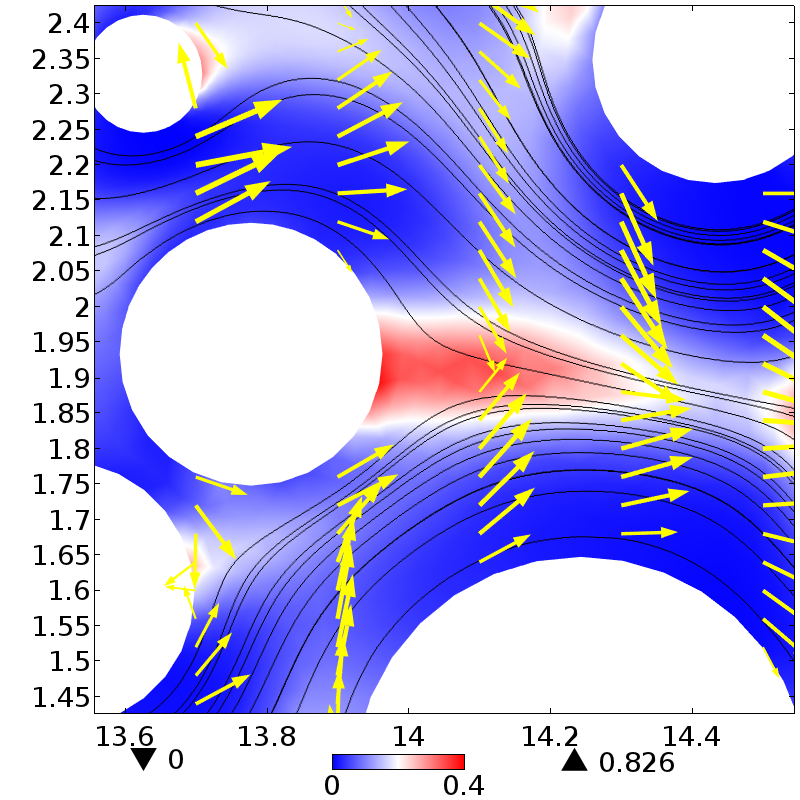}}\hfil
  \subfloat[]{\includegraphics[width=0.5\linewidth]{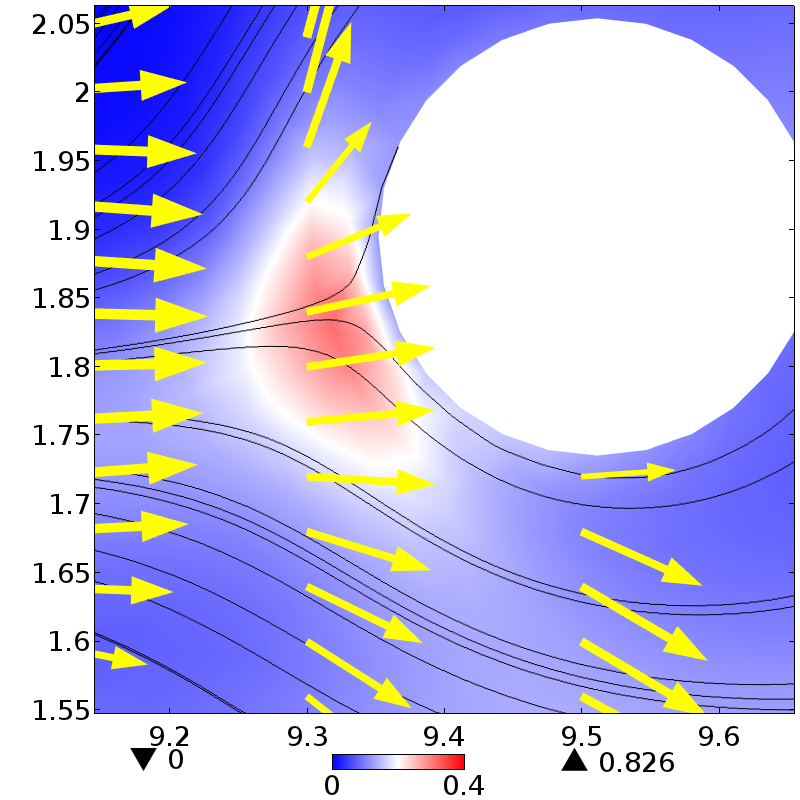}}

  \subfloat[]{\includegraphics[width=0.5\linewidth]{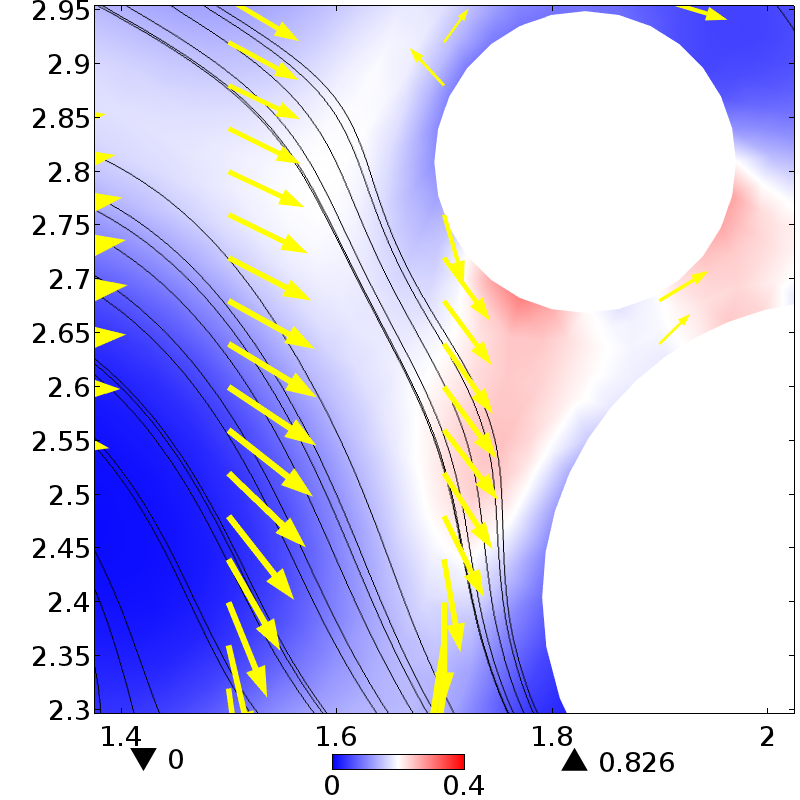}}\hfil
  \subfloat[]{\includegraphics[width=0.5\linewidth]{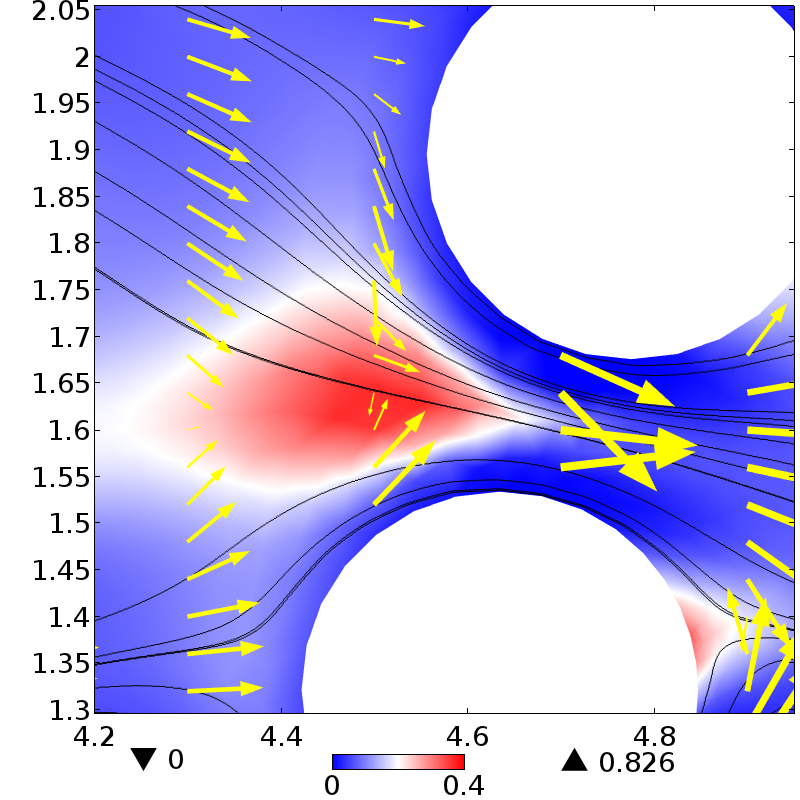}}\\
  \medskip
  \caption{Particle concentrations $\varphi$ (surface plot), water streamlines (solid black lines) and particles streamlines (yellow arrows) for four different regions of the domain: (a) behind a solid grain, (b) at a stagnation point, (c) in a no-flow region and (d) at the entrance of a narrow pore throat. Red colour indicate intense accumulation of particles.}
  \label{figEx.zoom}
\end{figure}

Next, particles are introduced, by injection with $\varphi=0.1$.
The resulting pattern of accumulation and dispersion is shown in the bottom 
of Fig. \ref{fig.divV}. 
Large rates of strain are caused by an abrupt change in the water velocity.
In this scenario, one can recognize four different characteristic regions where accumulation is favoured: in the wake behind solid grains, at stagnation points, in no-flow regions near high-velocity regions and at the entrance of narrow pore throats.
In Fig. \ref{figEx.zoom} we show four zoomed-in regions from Fig. \ref{fig.divV}, each with a different characteristic pattern of particle accumulation.
We also plot the water streamlines as solid black lines and particles velocity as yellow arrows.
Note that due to inertia, water and particles trajectories are not parallel.

Of the four accumulation scenarios, only accumulation at the pore entrance, 
Fig. \ref{figEx.zoom}(d), can significantly alter the water flow.
While in the other three cases the particles accumulate in low-velocity regions, at the pore entrance accumulation occurs in a high-velocity region.
Therefore, if accumulation is large at the pore entrance, the channel may experience a partial or total clogging, which causes a local redistribution of the pressure in the upstream side of the flow.
If the clogging is intense enough, the local pressure redistribution can divert the upstream water flow into neighbouring channels.

\begin{figure}[h!]
\centering
\includegraphics[width=1\linewidth]{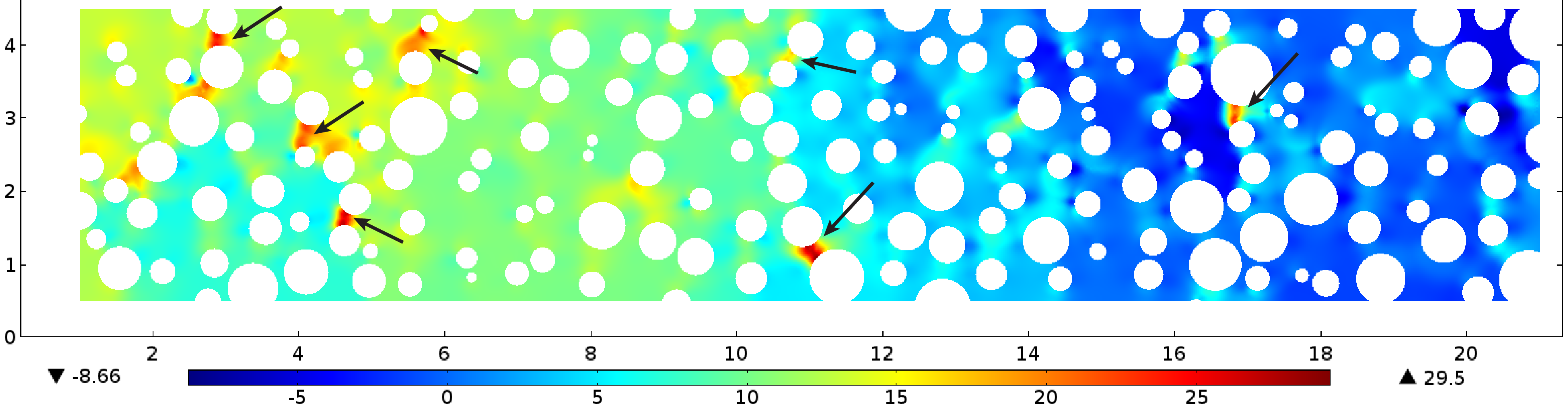}
\caption{Pressure difference between the flow without and with particles, $p_{np}-p_p$. Positive, red regions means that the pressure decreased, whereas negative, blue regions means that the pressure increased. The arrows point existing clogs.}
\label{figEx.pDiff}
\end{figure}

\begin{figure}[h!]
\centering
  \subfloat[]{\includegraphics[width=0.5\linewidth]{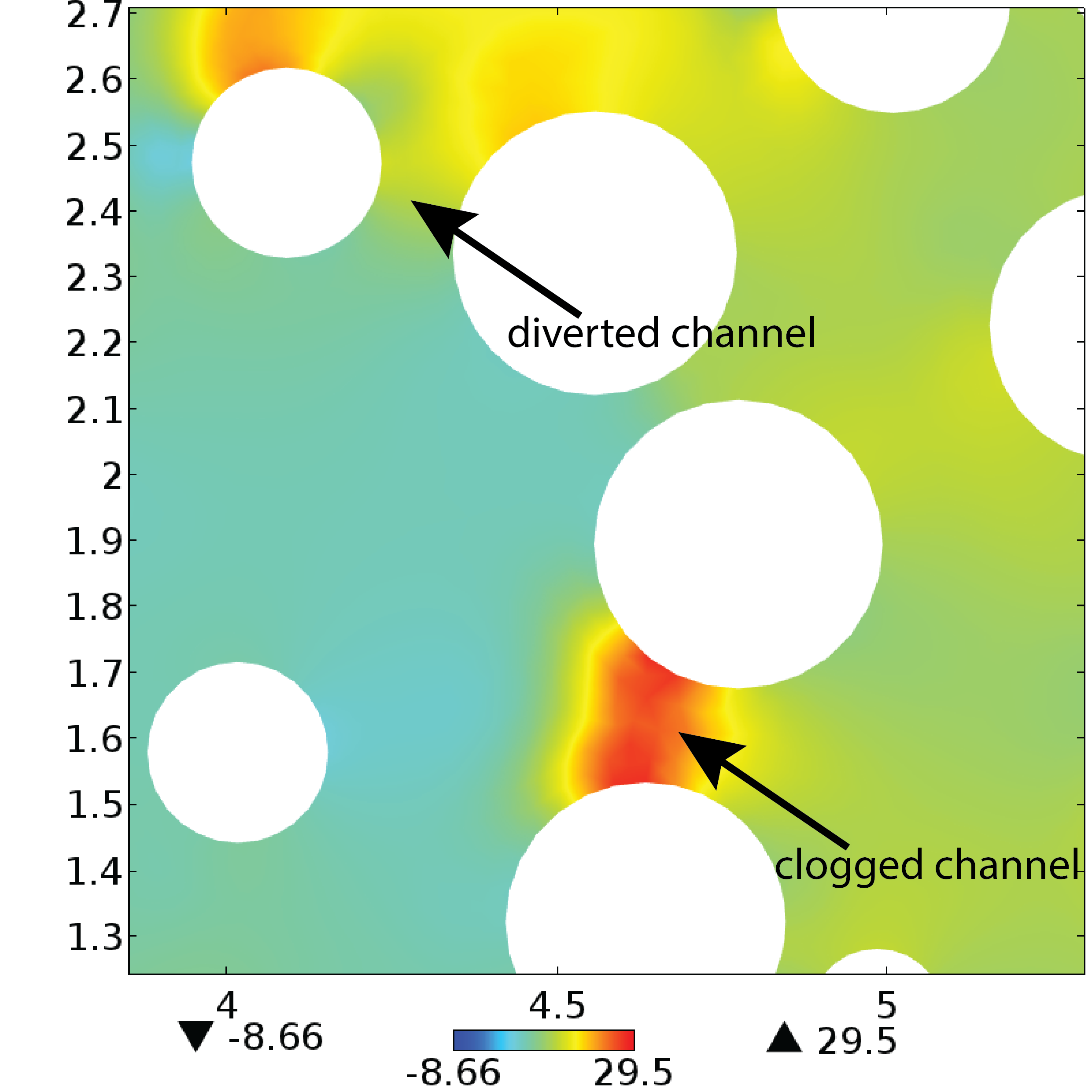}}\hfil
  \subfloat[]{\includegraphics[width=0.5\linewidth]{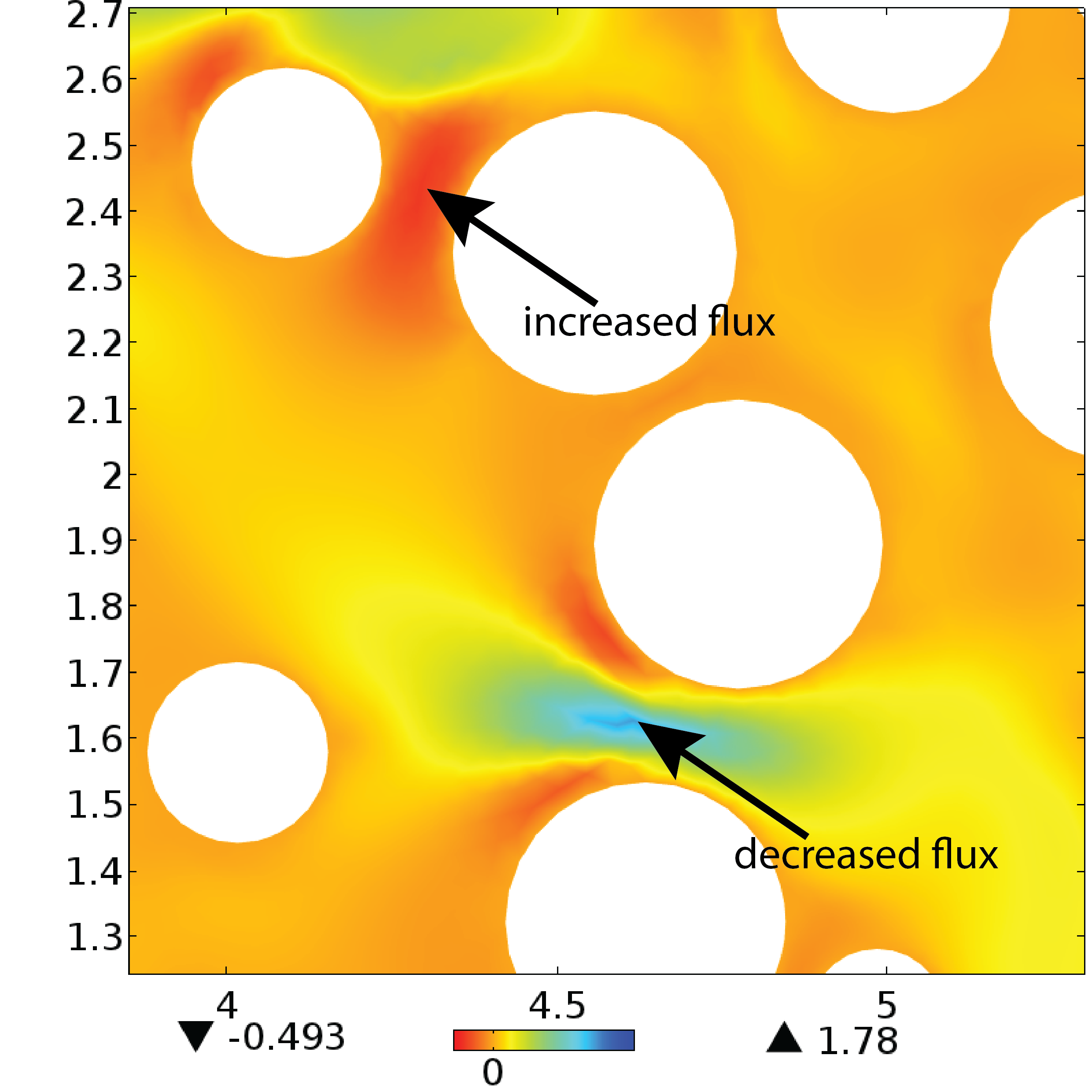}}\\
  \medskip
  \caption{Surface plot of the (a) pressure and (b) water flux differences between the cases without and with particles near a partially clogged channel.}
  \label{figEx.diffZOOM}
\end{figure}

In Fig. \ref{figEx.pDiff} we show $p_{np}-p_p$, the pressure difference between the flow without particles and the flow with particles.
The arrows in the domain point to existing clogs.
One can see that in some regions of intense accumulation at the pore-entrance, the pressure decreased (larger positive values in Fig. \ref{figEx.pDiff}), thereby causing the local pressure behind the channel to increase (negative values in Fig. \ref{figEx.pDiff}).
This is more clearly seen in Fig. \ref{figEx.diffZOOM}, where we show a zoomed-in region with the pressure difference and the difference between water fluxes divided by the average water flux in the case without particles, i.e.,
$(\Psi_{np} - \Psi_{p})/\overline{\Psi}_{np}$, where $\Psi = (1-\varphi)|\bm{u}|$ and $\overline{\Psi} = \Omega^{-1}\int_\Omega\Psi d\Omega$ is the average water flux in the domain $\Omega$.
Note that at the entrance of the partially clogged channel, there is a pressure decrease.
This happens because of the particle accumulation at the pore entrance, which decreases the water flow in the channel.
Total clogging is not observed because in our model we do not consider particle attachment at the wall.
Therefore, the region near the solid grain, which is dominated by vorticity, favours particle dispersion.

\begin{figure}[h!]
\centering
  {\includegraphics[width=0.6\linewidth]{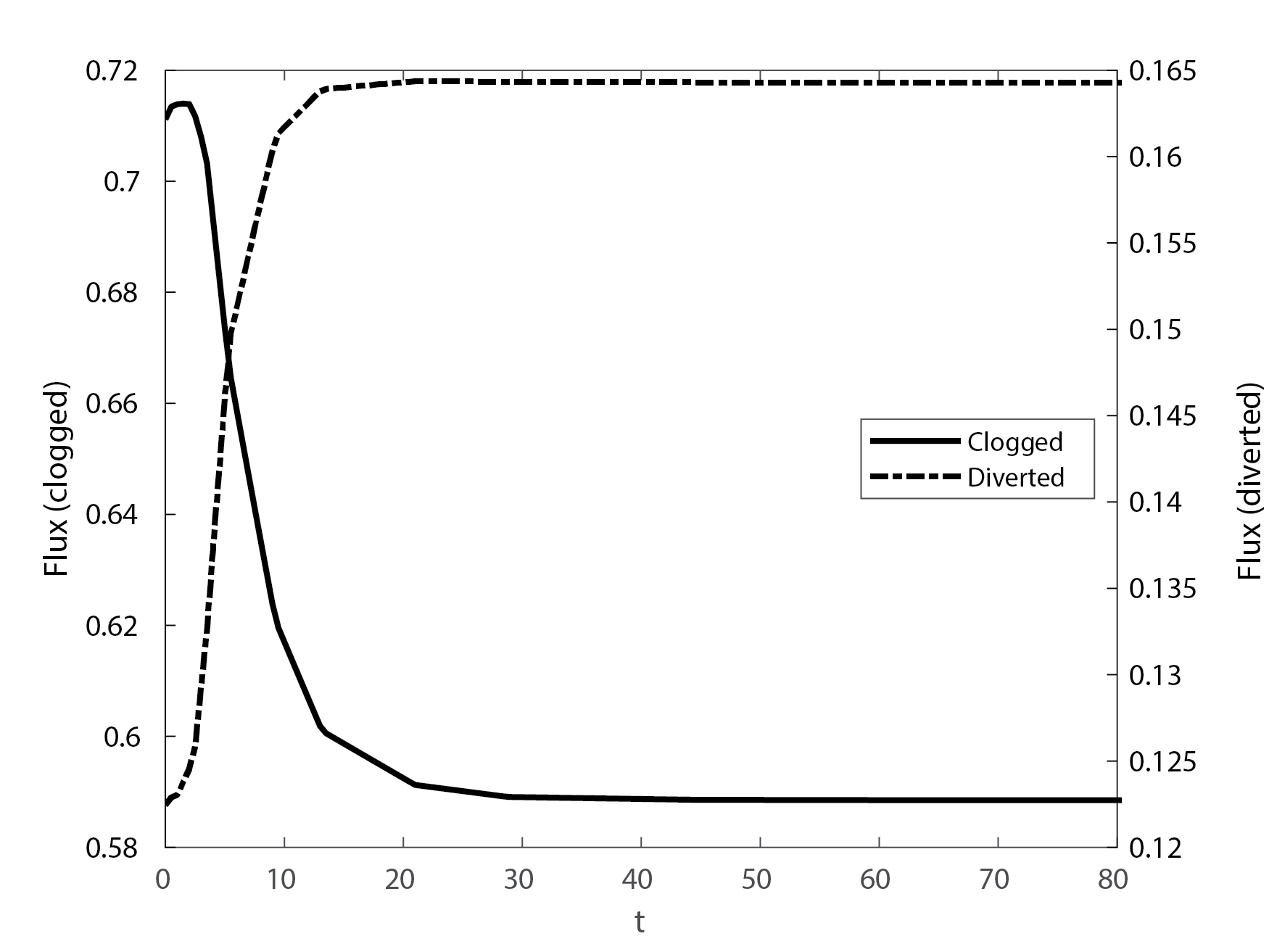}}
  \caption{Time variation of the water flux $\Psi = (1-\varphi)|\bm{u}|$ in the clogged and a nearby pore. Corresponding channels are indicated in Fig. \ref{figEx.diffZOOM}(a).}
  \label{figEx.flux}
\end{figure}

The occurrence of a clog can be identified by a pore throat presenting a 
significant decrease in the pressure at its entrance and a consequent decrease 
in its water flux.
Since clogging is usually not complete, the near-wall regions of the partially clogged pore experience a faster flow, as discussed previously.
Nevertheless, the overall water flux in the pore decreases, as we can see in Fig. \ref{figEx.flux}, where we  show the time variation of the water flux for the partially clogged channel and a nearby channel, both indicated in Fig. \ref{figEx.diffZOOM}.
Note that the decrease of the water flux in the partially clogged channel occurs at the same time scale as the increase in the water flux in the diverted channel.
This indicates that these changes are correlated.

\subsection{Impact of pore throat heterogeneity}

\begin{figure}[h!]
\centering
\includegraphics[width=0.65\linewidth]{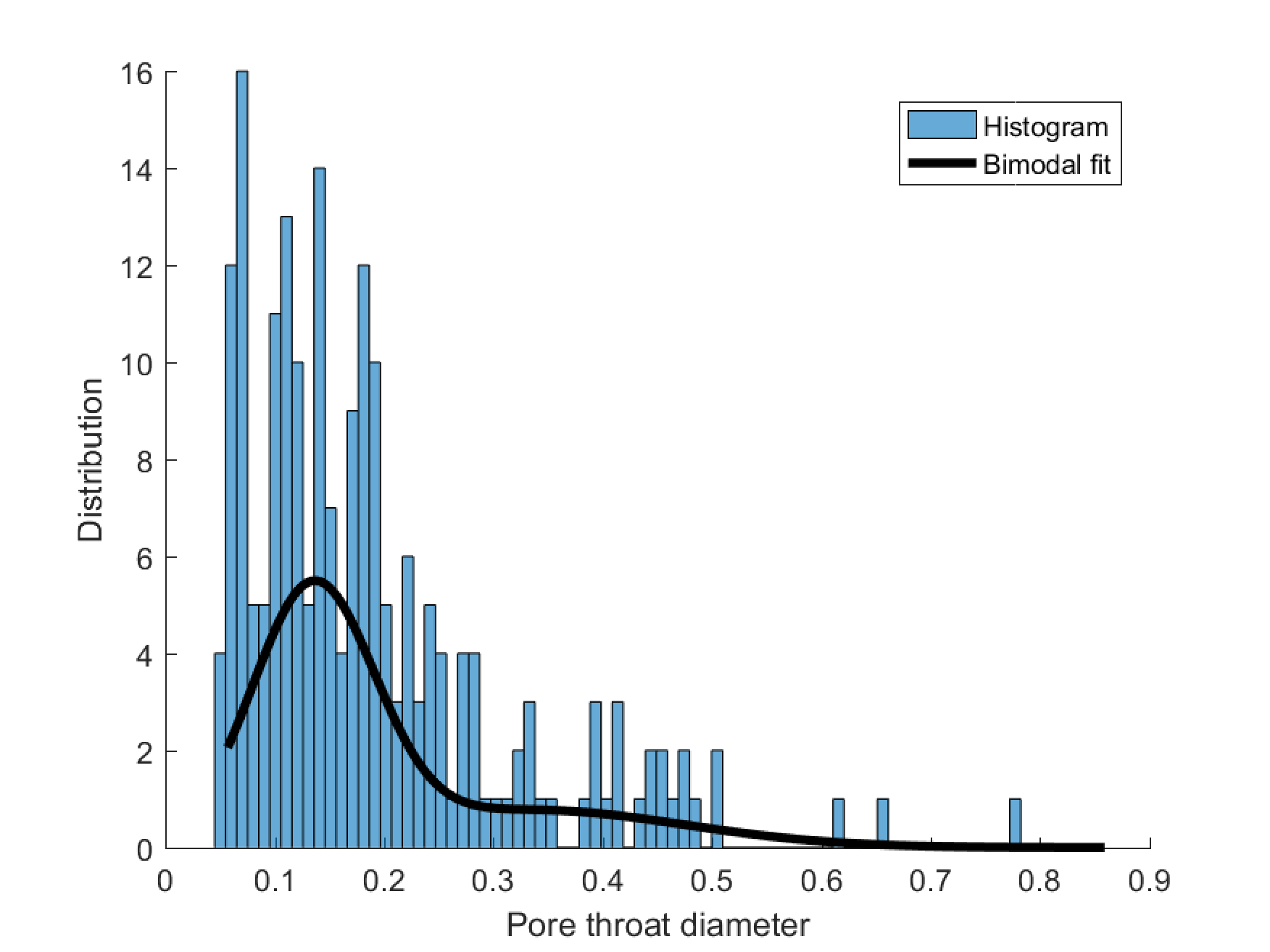}
\caption{Pore throat diameter distribution of the homogeneous medium and its bimodal fit.}
\label{figHom.PTDD}
\end{figure}

\begin{figure}[h!]
\centering
\includegraphics[width=1\linewidth]{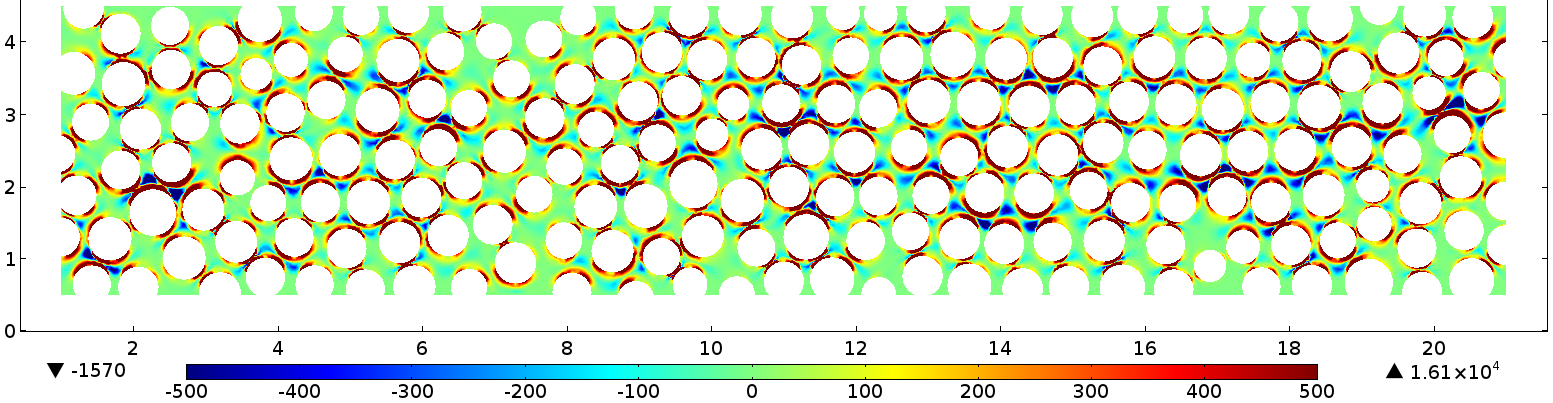}
\includegraphics[width=1\linewidth]{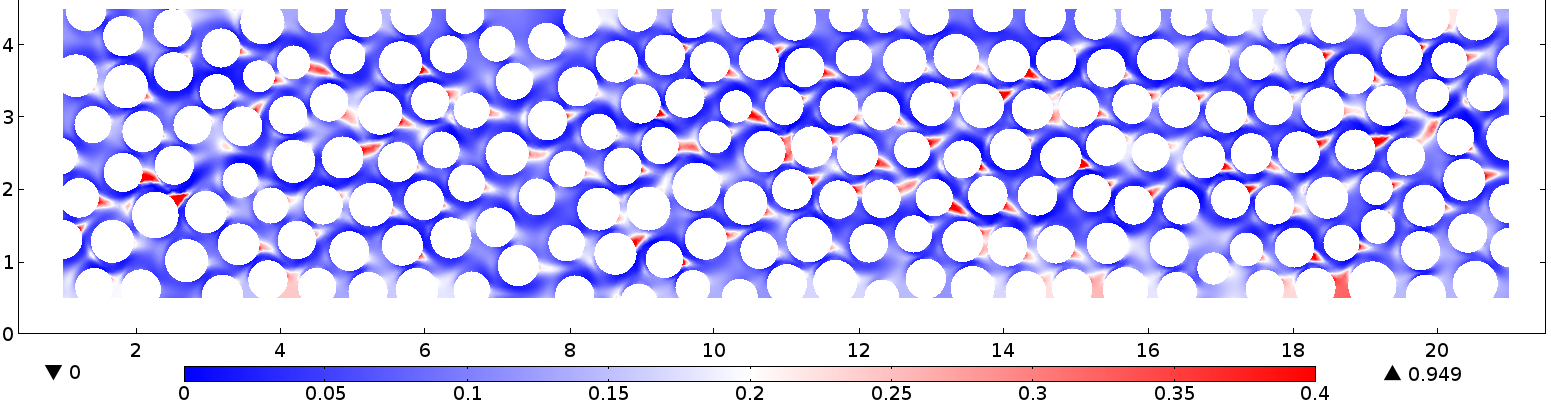}
\caption{Top part: surface plot of $\delta$. Bottom part: particles concentration for injection at $\varphi=0.1$.}
\label{figHom.divV}
\end{figure}

Since accumulation in a high velocity region is an effect of flow acceleration when going from a large to a narrow pore, it is expected that in a homogeneous medium this accumulation pattern is minimal, as most of the pores have the same size. 
To test this hypothesis, we consider a flow through the homogeneous medium shown in Fig. \ref{figHom.divV}.
For the initial edge lengths distribution, we use the same values shown in Tab. \ref{tab.parGumbel}, only changing $\alpha_1 = 0.7$.
This change is done to obtain a more homogeneous distribution of edges.
After we construct the solid grains, the homogeneous medium have a PTDD shown in Fig. \ref{figHom.PTDD}.
We fit the PTDD through a bimodal distribution, with parameters given in Tab. \ref{tab.hom}.
\begin{table}[h!]
\centering
\begin{tabular}{|c|c|}
\hline
 $p = 0.7289$         & $\gamma   = 0.9235$\\
 $\eta_1    = 0.1350$ & $\eta_2   = 0.3352$\\
 $\sigma_1 = 0.0557$  & $\sigma_2 = 0.1400$\\
\hline
\end{tabular}
\caption{Fitting parameters for the homogeneous medium shown in Fig. \ref{figHom.PTDD}}
\label{tab.hom}
\end{table}

\begin{figure}[h!]
\centering
\includegraphics[width=1\linewidth]{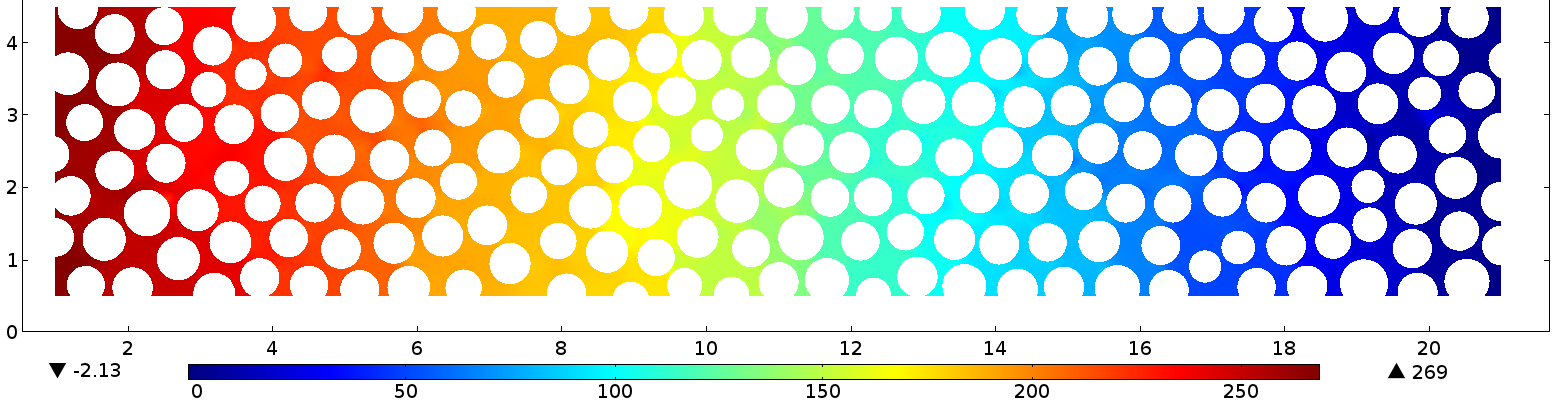}
\caption{Pressure difference between the cases without and with particles. The linear decrease in the pressure is caused by the presence of particles in the flow and is not associated with partial blockage of pore channels.}
\label{figHom.p}
\end{figure}

As discussed previously, it is expected that only significant accumulation in low-velocity regions will occur.
Here, we consider the same parameters as before ($Re = 5, St = 0.1$ and $\tilde{\rho} = 10$), but now with injection of water occurring at $u_{inj} = 0.6$.
This modification on the injection condition is done in order to have values of $\delta$ at the same order of magnitude as in the heterogeneous case, see upper part of Fig. \ref{figHom.divV}.
In Fig. \ref{figHom.divV} we show surface plots for $\delta$ and $\varphi$.
Note that we still have regions of predominant strain, $\delta<0$, and vorticity, $\delta>0$.
However, differently from what is observed for the heterogeneous medium (see Fig \ref{fig.divV}), the regions of dominating strain are mostly located in low-velocity regions.
Therefore, intense accumulation of particles occurs only in low-velocity regions, such that the flow pattern is not strongly modified by the accumulation of particles.
This feature can be seen in Fig. \ref{figHom.p}, where, as before, we plot the pressure difference for the cases without and with particles.

 {
The lack of accumulation in high-velocity regions means that blockage of pores, 
with a consequent pressure redistribution in the upstream side and possible 
flow diversion, is much less likely than in the heterogeneous case presented in 
the previous section.
This is expected, as the heterogeneous media present more paths connecting 
large to narrow pores, which is a key ingredient to accumulation in 
high-velocity 
regions, when compared to homogeneous media.
}
In the context of enhanced oil recovery through polymer particles injection, experimental results indicate that heterogeneous cores show increased rates of recovery compared to homogeneous cores \cite{spildo2009}.
These results are consistent with the proposed mechanism of particle accumulation and flow diversion leading to EOR.

Note that the predominance of same-size pores will exist no matter which homogeneous medium we consider.
Therefore, the lack of significant accumulation in high-velocity regions is a feature of every homogeneous media, such that the conclusions drawn in this section are quite general.
Formally, from our definition, homogeneous media are characterized by $\gamma\approx0$ and  $\gamma\gg1$.
 {
It is important to point, however, that not only the pore throats distribution is relevant to the formation of clogs, but connectivity also plays a significant role.
If the connection between a large and a narrow pore is not aligned with the flow direction, accumulation will occur in a low-velocity region.
For example, the accumulation pattern shown in Fig. \ref{figEx.zoom}(c) occurs in a narrow pore connected to a large pore.
However, since the connection between the large and the narrow pore is perpendicular to the flow direction, accumulation occurs in a low-velocity region, such that the flow structure is not affected.
Therefore, the rate of clog formation present in a heterogeneous cores depends on the connectivity between large and narrow pores.
A proper quantification of this requires establishing a link between pore scale geometry, connectivity and pore scale flow.
Nevertheless, our results serve to highlight the difference between the accumulation patterns that arise in heterogeneous and homogeneous media.
}

\subsection{Influence of $u_{inj}, Re$ and $\tilde{\rho}$}

\begin{figure}[h!]
\centering
  \subfloat[]{\includegraphics[width=0.5\linewidth]{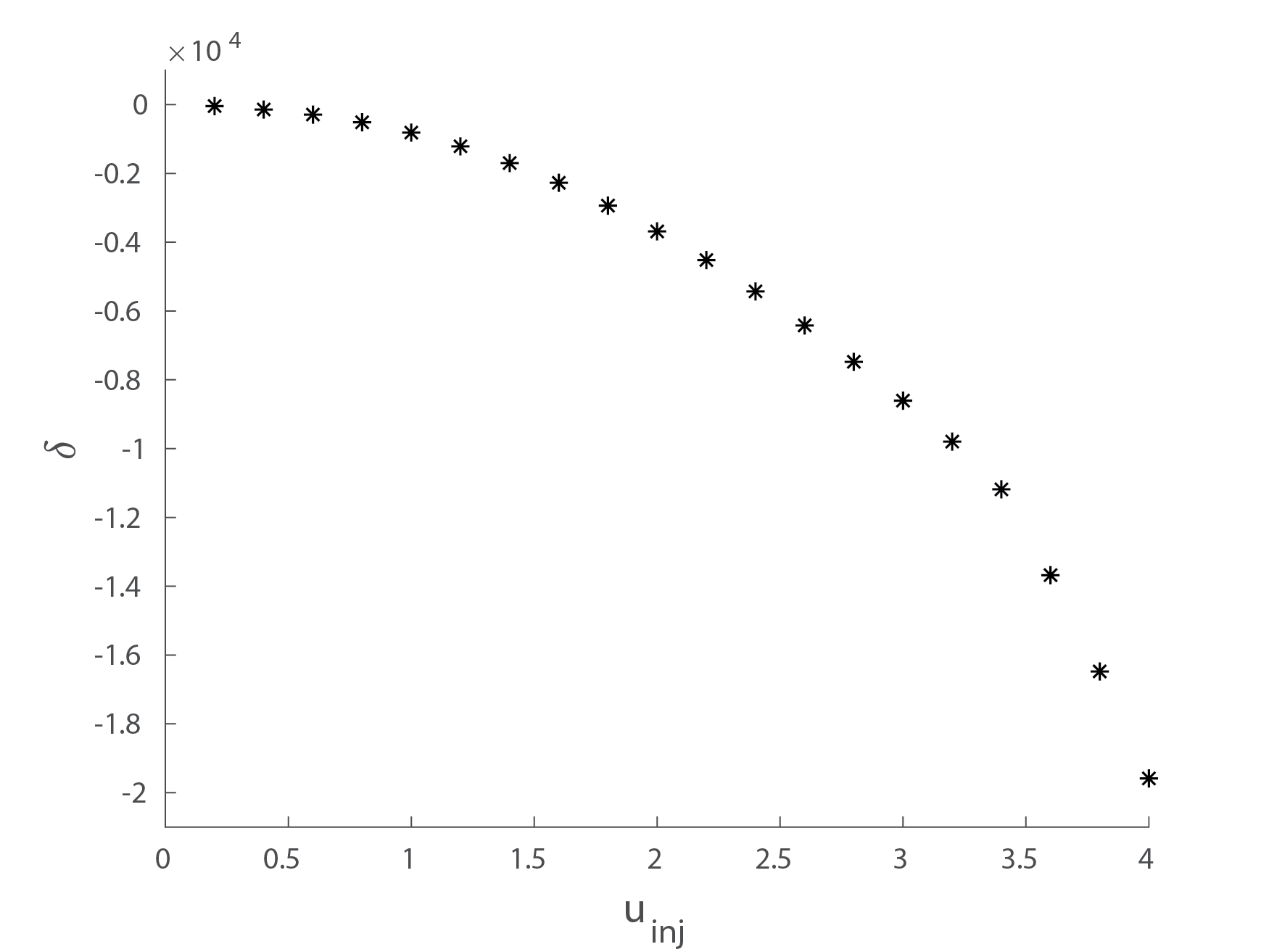}}\hfil
  \subfloat[]{\includegraphics[width=0.5\linewidth]{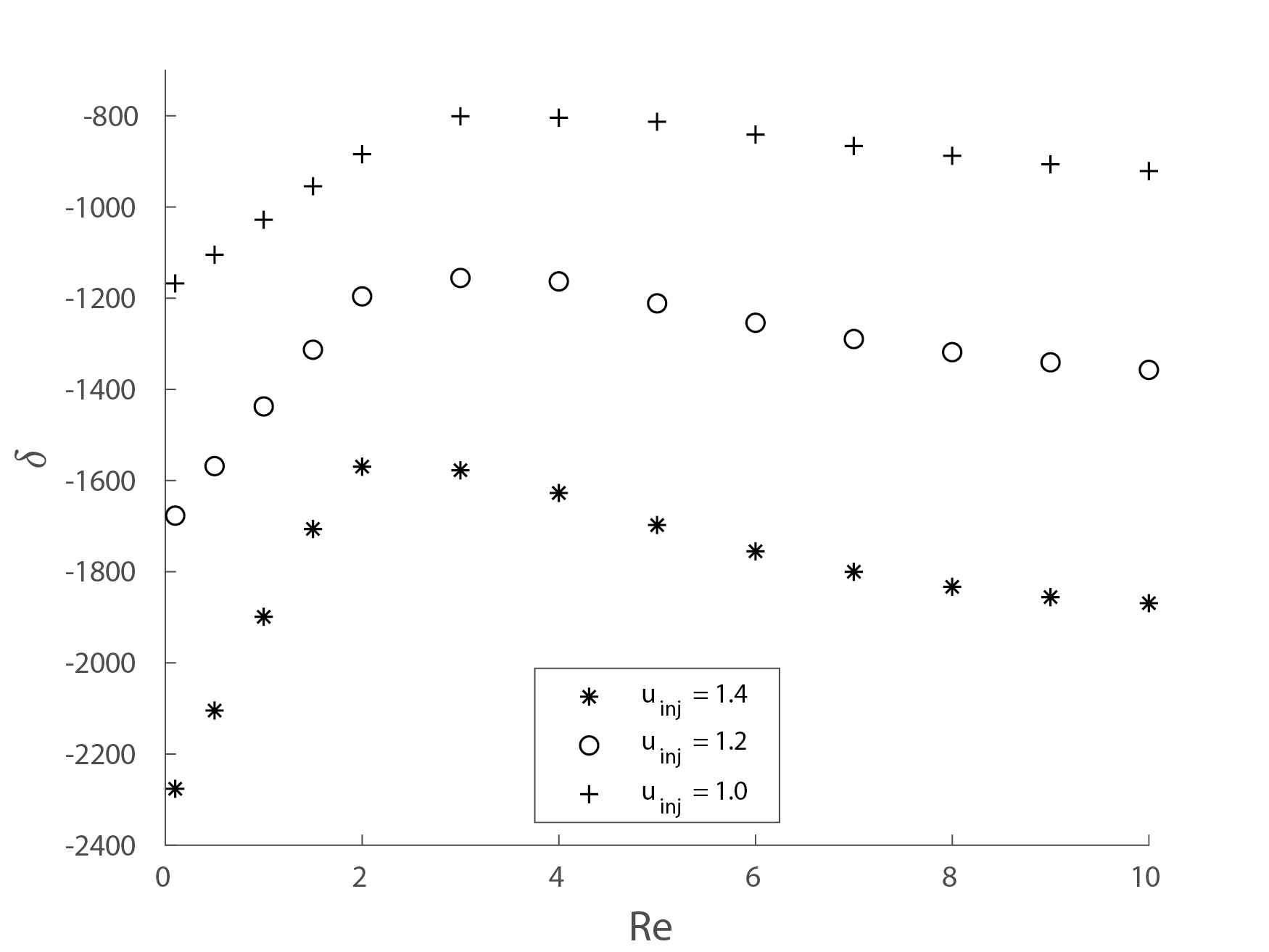}}\\
  \medskip
  \caption{(a) $\delta_{min}\times u_{inj}$ for $Re = 5$ and (b) $\delta_{min}\times Re$ for $u_{inj} = \{1.0, 1.2, 1.4\}$. Increasing values of the injection velocity leads to a decrease on $\delta_{min}$ and, consequently, on the accumulation rate. The variation with $Re$ presents two characteristic regimes: a Stokes regime (left branch) and a convection-dominated regime (right branch).}
  \label{fig.par}
\end{figure}

As discussed in the previous section, accumulation and dispersion patterns can be analysed through the quantity $\delta = St^{-1}\nabla\cdot\bm{v}$.
It is possible to evaluate $\delta$ in a flow scenario without particles in order to have an approximate relation between strain-dominated regions $\delta<0$ and accumulation, and between vorticity-dominated regions $\delta>0$ and dispersion.
In our approach, the maximum accumulation will occur in the region where $\delta$ achieves its minimal value.
Therefore, we can study the accumulation behaviour by analyzing the variation in $\delta_{min}$ with the problem parameters.
We do so by considering the heterogeneous medium shown in Fig. \ref{fig.porScheme}.

For increasing values of the injection velocity $u_{inj}$, the rates of strain and vorticity in the medium increase.
This leads to a decrease in $\delta_{min}$, as seen in Fig. \ref{fig.par}(a), where we plot $\delta_{min}$ for a fixed value of $Re = 5$.
Therefore, for increasing injection velocities it is expected an increase in accumulation.
For a fixed value of $u_{inj}$, the variation of $\delta_{min}$ with $Re$ is non-monotonous, as shown in Fig. \ref{fig.par}(b), where we consider $u_{inj} = \{1.0; 1.2; 1.4\}$.
For small values of $Re$, an increase in the Reynolds number leads to an increase in $\delta_{min}$, thus lowering accumulation of particles.
However, after a critical value of $Re$, an increase in the Reynolds number leads to a decrease on $\delta_{min}$, thus enhancing accumulation of particles.
An increase in $Re$ means a decrease on the thickness of the viscous boundary layer at the solid grains surface.
Since the injection velocity $u_{inj}$ is constant, an increase in $Re$ means that the overall flow velocity increases.
Therefore, for the left branch, an increase in the velocity leads to a decrease on the accumulation of particles, whereas for the right branch it leads to an increase on the accumulation.
We refer to the left branch as Stokes regime and to the right branch as convection-dominated regime.
 {
Since $\bm{v} \sim -\bm{u}\cdot\nabla\cdot\bm{u}$ and $\delta_{min} \sim -\nabla\cdot(\bm{u}\cdot\nabla\bm{u})$, accumulation and dispersion of particles are generally favoured by an increase in velocity.
This is the case for the convective-dominated regime.
However, in the Stokes regime the convective transport is of higher-order with respect to the viscous and pressure forces, such that the left-hand side of Eq. (\ref{eqAL.momW}) is very small in comparison to the right-hand side.
Thus, when $Re$ is small the velocity of particles approximates as
}
\begin{equation}
 \bm{v}
 =
 \bm{u} - St\left(\frac{1}{\tilde{\rho}Re}\nabla p + \frac{\partial\bm{u}}{\partial t} + \bm{u}\cdot\nabla\bm{u}\right)
 \approx
 \bm{u} - \frac{St}{\tilde{\rho}Re}\nabla p,
 \label{eqRe.v}
\end{equation}
 {
such that $\nabla\cdot\bm{v} \approx \nabla\cdot\bm{u}$.
Thus, in the Stokes regime an increase on the overall flow velocity does not enhance accumulation.
Rather, it decreases it as particles are transported into the downstream direction.
}
As shown in Fig. \ref{fig.par}(b), the Stokes regime shrinks for increasing injection velocities, as expected.

\begin{figure}[h!]
\centering
  \subfloat[]{\includegraphics[width=0.9\linewidth]{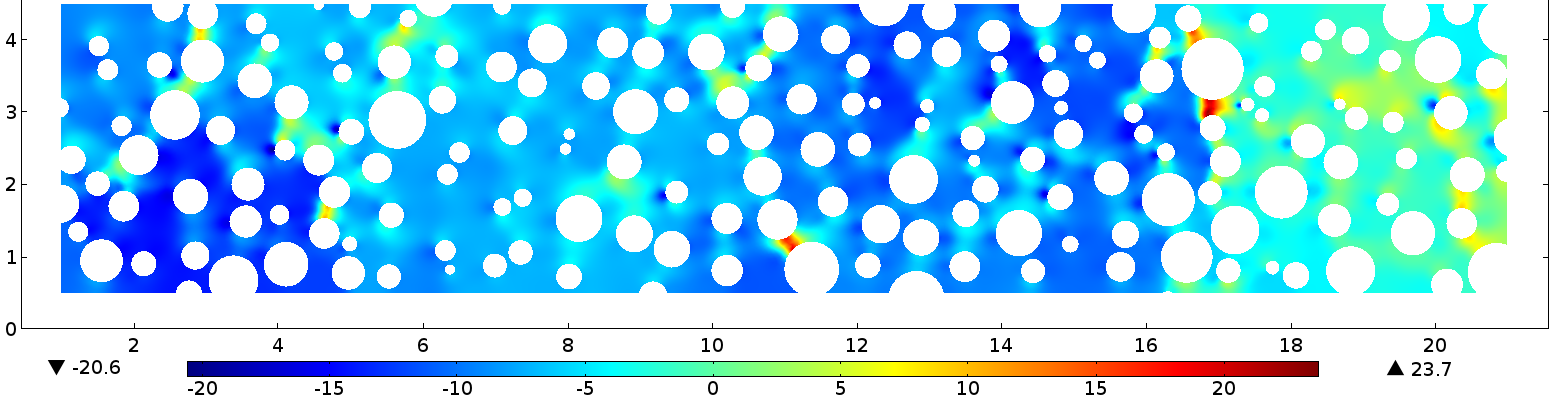}}\\
  \subfloat[]{\includegraphics[width=0.9\linewidth]{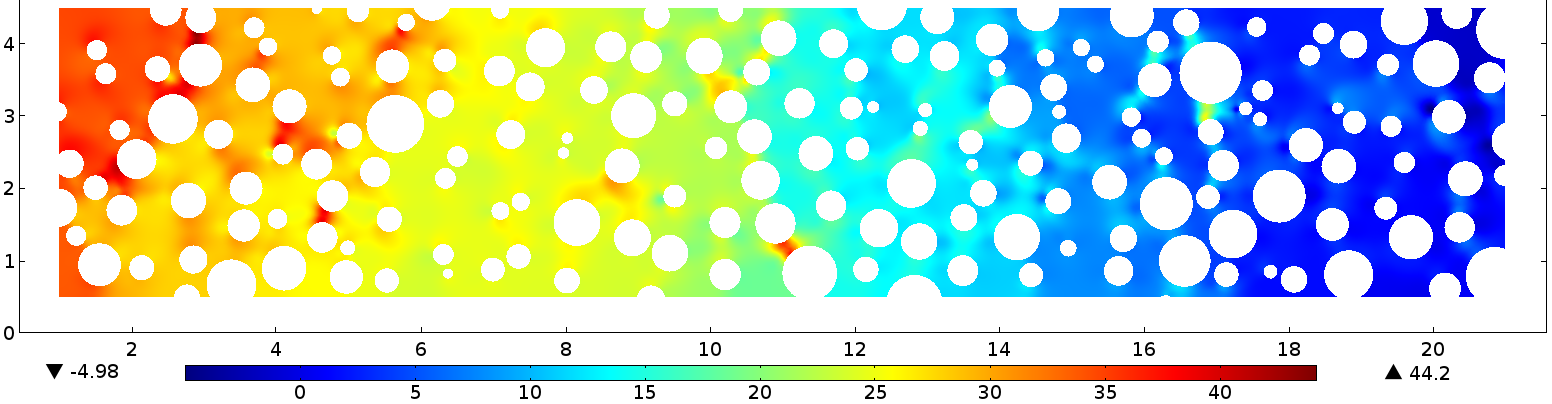}}\\
  \subfloat[]{\includegraphics[width=0.9\linewidth]{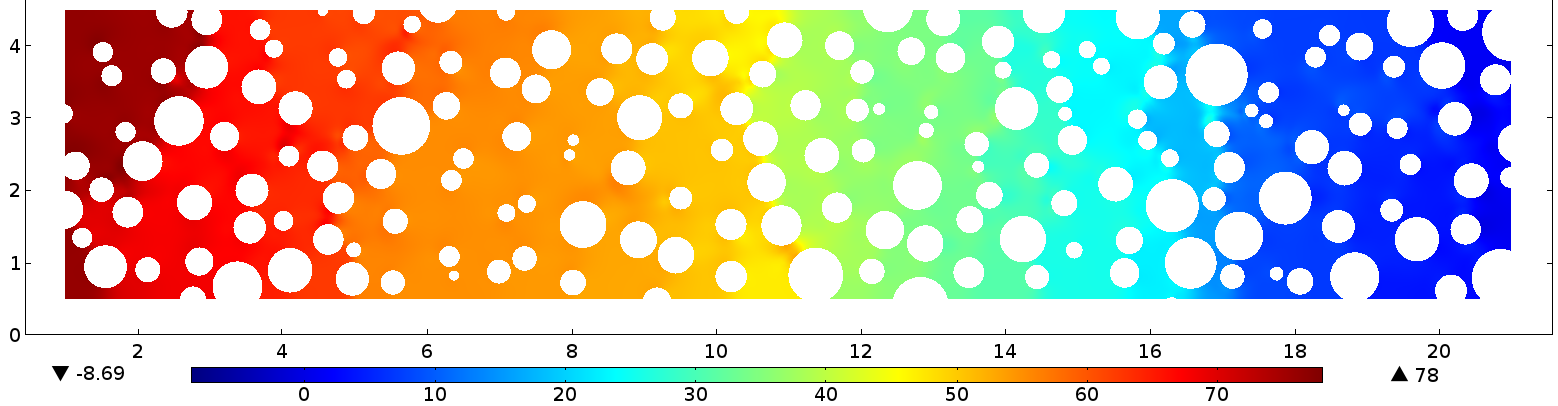}}\\
  \medskip
  \caption{Pressure difference for (a) $\tilde{\rho} = 15$, (b) $\tilde{\rho} = 10$ and (c) $\tilde{\rho} = 5$. Decreasing values of $\tilde{\rho}$ increase the particles velocity $\bm{v}$ and favour the disappearance of clogs (recalling that clogs are characterized by an intense decrease on the pressure at the pore entrance).}
  \label{fig.rho}
\end{figure}

As shown by (\ref{eq.divVphi0B}), strain-dominated regions favour accumulation.
However, as pointed out by Eq. (\ref{eqAL.massPb}), this occurs along the streamlines of the particles.
If the particles velocity increases considerably, accumulation locations move further downstream inside the domain.
For accumulation in low velocity regions, this poses a minor difference, as the particles streamlines are barely modified in such regions.
On the other hand, for accumulation in high velocity regions, this favours dispersion of particles, as they tend to move further downstream, thus lowering local accumulation.
In Fig. \ref{fig.rho} we plot the pressure difference for illustrative values of $\tilde{\rho} = 15, 10, 5$.
A decrease in the particle-to-water mass densities ratio leads to an increase in the particles velocity, according to Eq. (\ref{eqAL.momP}).
We see that the formation of clogs is unfavoured for decreasing values of $\tilde{\rho}$ (recalling that a clog is characterized by a significant decrease on the pressure at the pore entrance).
Writing Eq. (\ref{eqAL.massPb}) in Eulerian coordinates yields
\begin{equation}
 \frac{1}{\varphi}\frac{\partial\varphi}{\partial t}
 =
 -\nabla\cdot\bm{v} - \frac{1}{\varphi}\bm{v}\cdot\nabla\varphi.
 \label{eq.partEuler}
\end{equation}
Then, if we evaluate the local variation of $\varphi$, we see that particles with higher velocities tend to disperse accumulation due to the convective transport (second term in the right-hand side of Eq. (\ref{eq.partEuler})).
This makes clogs dissipate if the particles velocity is high enough.
This is seen on the surface plot of the pressure differences, Fig. \ref{fig.rho}.

\section{Conclusions}

 {
We propose that accumulation of particles transported by water in porous media can be explained by hydrodynamic effects.
In order to test our hypothesis, we consider a simple multiphase flow model for the transport of inertial particles by water at the pore-scale.}
The tortuous paths of the random medium generate regions of dominating strain and vorticity, which favour accumulation and dispersion of particles, respectively.
The numerical results show that heterogeneous media present significant accumulation in low and high-velocity regions, whereas homogeneous media present only significant accumulation in low-velocity regions.
 {
Accumulation in a high-velocity region leads to the formation of a clog, which causes a redistribution of the pressure on the upstream side and consequently to flow diversion.
Thus, clog formation and microscopic flow diversion occur predominantly in heterogeneous media.
%
%
}

Detailed knowledge of the water flow pattern in the porous medium without particles can provide information as to where accumulation is favoured by identifying strain-dominated regions.
Near the surface of the solid grains vorticity dominates and particles tend to disperse from it.
 {
The model presented in this paper can be upscaled to analyse inertia effects at the Darcy scale.
Usually, upscaled models for particle deposition consider that the particles are advected with the same velocity as the water \cite{gruesbeck1982,jegatheesan2005}.
Thus, inertia effects are not captured in conventional upscaled models.
Moreover, incorporating the effects of pore-scale geometry into the upscaled equations is a challenge.
}

As particles begin to accumulate, interparticle interaction becomes relevant.
In such scenario, the specification of the type of particles under consideration is crucial.
 {
In particular, it is known that polymer particles form gels at increasing concentrations \cite{bjorsvik2008,marliere2015}.
Therefore, accumulation of particles at the entrance of a narrow pore throat can trigger the formation of stronger gels, which can fully clog a channel.
}
Extending the model to account for the interactions of the particles requires specification of particle rheology.
%

\section*{Acknowledgements}

MAEK, EK and KS are supported by Equinor through the Akademia agreement. The work of AM is partially supported via MultiBarr (KK Foundation, project nr. 20180036). FAR was partially supported by the Norwegian Academy of Science and Letters and Equinor through the VISTA AdaSim project $\#6367$.  ISP was supported by the Research Foundation-Flanders (FWO) through the Odysseus programme (project GOG1316N) and by Equinor through the Akademia agreement. KK work was partially funded by the Research Council of Norway (NFR) Projects 811716 LAB2FIELD and 810857 MICAP.

\appendix

\section{Model formulation}

The Eulerian description of the multiphase flow model is obtained by averaging local quantities and considering immiscible phases \cite{bowenBook,gidaspowBook,pekerBook}.
The general momentum transport for each phase accounts for the momentum exchange between phases.
If $\varphi_i,\rho_i$ and $\bm{u}_i$,  represent the volumetric occupation, mass density and velocity of phase $i$, respectively, the general conservation equations without external forces can be stated as
\begin{align}
\frac{\partial}{\partial t}(\rho_i\varphi_i)
+
\nabla\cdot(\rho_i\varphi_i\bm{u}_i)
&= 0,
\label{eqApp.Mass}
\\[5pt]
\frac{\partial}{\partial t}(\rho_i\varphi_i\bm{u}_i)
+
\nabla\cdot(\rho_i\varphi_i\bm{u}_i\bm{u}_i)
&=
\nabla\cdot\bm{T}_i + \bm{F}_{i,I} + \bm{f}_i,
\label{eqApp.Mom}
\end{align}
where $\bm{T}_i$ is the stress tensor, $\bm{F}_{i,I}$ the forces acting at the interface of the phases and $\bm{f}_i$ is the source of momentum due to the interaction between phases.
For a closed system of $N$ components, it is required that $\sum_{i=1}^N\bm{f}_i = \sum_{i=1}^N\bm{F}_i = 0$ and $\sum_{i=1}^N\varphi_i = 1$.
In order to close the formulation, one needs to specify the interaction terms $\bm{f}_i$, the interface forces $\bm{F}_{i,I}$ and the constitutive expressions for $\bm{T}_i$.

We consider a solid-liquid system, such that $i=s,l$.
For simplicity, we consider only Stokes drag as the interaction force between phases.
Therefore, $\bm{f}_l = -\bm{f}_s = \phi_s\rho_s(\bm{u}_s - \bm{u}_l)/t_s$, see Jackson \cite{jackson1996}.
The general expression for the stress tensor of a compressible phase is given by \cite{bowenBook,gidaspowBook}
\begin{equation}
\bm{T}_i = -\varphi_i p_i\bm{I} + \varphi_i\mu_i(\nabla\bm{u}_i + \nabla\bm{u}_i^T) - \frac{2}{3}\varphi_i\mu_i\bm{I}\nabla\cdot\bm{u}_i.
\label{eqApp.ti}
\end{equation}
The viscosity of the solid particles $\mu_s$ is a measurement of the exchange of momentum between solid particles in adjacent layers.
If no collisions occur, we have $\mu_s = 0$, such that for the solid phase the stress tensor is given by
\begin{equation}
\bm{T}_s = -\varphi_s p_s \bm{I}.
\label{eqApp.ts}
\end{equation}

The averaged balance of momentum at the interface states that (neglecting the forces due to surface tension) \cite{pekerBook}
\begin{equation}
\sum_i \bm{T}_{i,I}\cdot\nabla\varphi_i = \sum_i \bm{F}_{i,I}\cdot\nabla\varphi_i = 0.
\label{eqApp.int}
\end{equation}
For each $i$ component, we have $\bm{F}_{i,I} = \bm{T}_{i,I} = - p_{i,I}\bm{I} + \bm{\tau}_{i,I}$.
When we plug $\bm{F}_{i,I}$ in Eq. (\ref{eqApp.Mom}) with $i=l,s$, the following terms appear
\begin{equation}
(p_l-p_{l,I})\cdot\nabla\varphi_l, \ \ \ (\bm{\tau}_l - \bm{\tau}_{l,I})\cdot\nabla\varphi_l.
\label{eqApp.intB}
\end{equation}
If we consider small and rigid particles (no deformation), the pressure inside a particle must be equal to the pressure at the interface, i.e., $p_s = p_{s,I} = p_{l,I}$, where the last equality is valid due to the non-deformation of particles.
Therefore, the momentum balance at the interface (\ref{eqApp.int}) for the solid phase is
\begin{equation}
\bm{\tau}_{s,I}\cdot\nabla\varphi_s = 0,
\label{eqApp.intS}
\end{equation}
which implies $\bm{\tau}_{s,I} = 0$ and by continuity $\bm{\tau}_{l,I} =0$.
Thus, we arrive at the following system of governing equations
\begin{align}
\frac{\partial}{\partial t}(\rho_l\varphi_l)
+
\nabla\cdot(\varphi_l\bm{u}_l)
&=
0,
\label{eqApp.massW}
\\[5pt]
\frac{\partial}{\partial t}(\rho_s\varphi_s)
+
\nabla\cdot(\varphi_s\bm{u}_s)
&=
0,
\label{eqApp.massP}
\\[5pt]
\rho_l\varphi_l
\left(
\frac{\partial\bm{u}_l}{\partial t}
+
\bm{u}_l\cdot\nabla\bm{u}_l
\right)
&=
-
\varphi_l
\nabla p
+
\nabla\cdot
\left(
\varphi_l\bm{\tau}_l
\right)
-
\frac{\varphi_s\rho_s}{t_s}(\bm{u}_l-\bm{u}_s),
\label{eqApp.momW}
\\[5pt]
\rho_s\varphi_s
\left(
\frac{\partial\bm{u}_s}{\partial t}
+
\bm{u}_s\cdot\nabla\bm{u}_s
\right)
&=
-
\varphi_s\nabla p
+
\frac{\varphi_s\rho_s}{t_s}(\bm{u}_l-\bm{u}_s).
\label{eqApp.momP}
\end{align}

In summary, the assumption of no collisionless, small, rigid particles eliminates the necessity of evaluating the interface dynamics from our model.
To close, we chose $p_l = p_s = p$, such that the pressure gradient contributes to the movement of each phase, only weighted by its respective volumetric occupation.
This is physically consistent with disregarding interparticle interaction, as no extra pressure due to collisions will arise.

\bibliographystyle{ieeetr}

\bibliography{refs}

\end{document}